\documentclass[reprint,aps,prb,superscriptaddress,10pt]{revtex4-2} 
\usepackage{graphicx}
\usepackage{bm}
\usepackage{dcolumn}
\usepackage{amssymb}
\usepackage{amsmath}
\usepackage{amsbsy}
\usepackage{amsfonts}
\usepackage{epsfig}
\usepackage{graphicx}
\usepackage{multirow}
\usepackage{stackrel}
\usepackage{paralist}
\usepackage{braket}
\usepackage{titlesec}
\usepackage{placeins}
\usepackage{color}
\usepackage[rgb]{xcolor}

\usepackage[percent]{overpic}
\usepackage{hyperref}
\hypersetup{colorlinks=true,allcolors=blue}
\usepackage{tikz}
\usetikzlibrary{intersections} 
\usetikzlibrary{positioning,calc}
\usepackage[caption=false]{subfig}
\usepackage[normalem]{ulem}

\tikzset{
    right angle quadrant/.code={
        \pgfmathsetmacro\quadranta{{1,1,-1,-1}[#1-1]}     
        \pgfmathsetmacro\quadrantb{{1,-1,-1,1}[#1-1]}},
    right angle quadrant=1, 
    right angle length/.code={\def\rightanglelength{#1}},   
    right angle length= 1 ex, 
    right angle symbol/.style n args={3}{
        insert path={
            let \p0 = ($(#1)!(#3)!(#2)$) in     
                let \p1 =  ($(\p0)!\quadranta*\rightanglelength!(#3)$),
                \p2 = ($(\p0)!\quadranta*\rightanglelength!(#2)$) in 
                let \p3 = ($(\p1)+(\p2)-(\p0)$) in  
            (\p1) -- (\p3) -- (\p2)
        }
    }
}

\newcommand{\cancel}[1]{}

\definecolor{NewColor}{rgb}{0,0,1}
\definecolor{myRed}{rgb}{1,0,0}
\definecolor{myGreen}{rgb}{0.2,0.6,0.2}
\definecolor{myBlue}{rgb}{0,0,1}

\graphicspath{{../Graphics/}{../Graphics/Publication/}}

\newcommand{\CO}[1]{\textcolor{red}{}}

\newlength{\noLengthl}

\renewcommand{\vec}[1]{{\boldsymbol{#1}}}

\usepackage{mathtools}
\usepackage{pict2e}
\usepackage{xcolor}
\usepackage{colortbl}
\graphicspath{{./}{./Figures_latex/Groundstate_Spins/}{./Figures_latex/3x3_RotationSpins/}{./Figures_latex/Figure3/}{./Figures_latex/Figure2/}{./Figures_latex/Figure1/}}

\newcommand{\beginsupplement}{%
	\setcounter{table}{0}
	\renewcommand{\thetable}{A\arabic{table}}%
	\setcounter{figure}{0}
	\renewcommand{\thefigure}{A\arabic{figure}}%
	\setcounter{equation}{0}
	\renewcommand{\theequation}{A\arabic{equation}}%
}

\definecolor{NewPink}{rgb}{0.9,0.074,0.78}
\definecolor{NewOrange}{rgb}{0.92,0.635,0.063}

\begin{document}
	
\title{
	Controlled Creation of Quantum Skyrmions
}
\author{Pia Siegl}
\affiliation{I. Institut f{\"u}r Theoretische Physik, Universit{\"a}t Hamburg, Notkestra{\ss}e 9, 22607 Hamburg, Germany}
\affiliation{The Hamburg Centre for Ultrafast Imaging, Luruper Chaussee 149, 22761 Hamburg, Germany}
\author{Elena Y. Vedmedenko}
\affiliation{Fachbereich Physik, Universit{\"a}t Hamburg, Jungiusstra{\ss}e 9, 20355 Hamburg, Germany}
\author{Martin Stier}
\affiliation{I. Institut f{\"u}r Theoretische Physik, Universit{\"a}t Hamburg, Notkestra{\ss}e 9, 22607 Hamburg, Germany}
\author{Michael Thorwart}
\affiliation{I. Institut f{\"u}r Theoretische Physik, Universit{\"a}t Hamburg, Notkestra{\ss}e 9, 22607 Hamburg, Germany}
\affiliation{The Hamburg Centre for Ultrafast Imaging, Luruper Chaussee 149, 22761 Hamburg, Germany}
\author{Thore Posske}
\affiliation{I. Institut f{\"u}r Theoretische Physik, Universit{\"a}t Hamburg, Notkestra{\ss}e 9, 22607 Hamburg, Germany}
\affiliation{The Hamburg Centre for Ultrafast Imaging, Luruper Chaussee 149, 22761 Hamburg, Germany}
\begin{abstract}
	We study the creation of quantum skyrmions in quadratic nanoscopic lattices of quantum spins coupled by Dzyaloshinkii-Moriya and exchange interactions. We numerically show that different kinds of quantum skyrmions, characterized by the magnitude of their spin expectation values and strong differences in their stability, can appear as ground state and as metastable excitations. In dependence on the coupling strengths and the lattice size, the adiabatic rotation of magnetic control fields at the boundary allows for the creation of quantum skyrmions. 
\end{abstract}

\maketitle	

\section{Introduction}
Magnetic skyrmions \cite{Bogdanov1989} are twisted magnetic structures which attract broad research interest due to their extra-ordinary real-space topological properties \cite{Nagaosa2013} and their potential usage in data storage devices \cite{Kiselev2011,Fert2017,Back2020}.
Being characterized by non-trivial winding characteristics of the magnetization \cite{Nagaosa2013}, magnetic skyrmions show a remarkable stability against local perturbations \cite{Vedmedenko2019}. 
Skyrmion lattices have been experimentally realized in bulk magnetic systems such as MnSi \cite{Muhlbauer2009}, $\text{Fe}_{0.5}\text{Co}_{0.5}\text{Si}$ \cite{Yu2010} and FeGe \cite{Kotani2018} and at interfaces of, e.g., Fe/Ir(111) \cite{Romming2013,Hagemeister2016,Vedmedenko2020, Heinze2011}.
Furthermore, single magnetic skyrmions can be created as metastable excitations via the injection of currents \cite{Sampaio2013, Yuan2016, Stier2017}, magnetic field pulses \cite{Flovik2017} and by tailored  boundary conditions \cite{Schaffer2020,Raeliarijaona2018}.
Depending on the involved magnetic interactions, magnetic skyrmions range in size from $\sim1~\mu \text{m}$ in bulk systems to $\sim1~\text{nm}$ at interfaces \cite{Nagaosa2013,Romming2013,Heinze2011}. While topological protection yields stability of large magnetic skyrmions, smaller skyrmions can be more easily created or annihilated by overcoming a finite energy barrier \cite{Hagemeister2015,Siemens2016,CortesOrtuno2017}.
Also, quantum effects rise in importance for a decreasing system size and spin quantum number.
Considering quantum corrections, further stabilization and decay mechanisms influence the stability of a magnetic skyrmion \cite{RoldanMolina2015,DerrasChouk2018}. On the one hand,
quantum spin fluctuations lead to a zero-point energy that stabilizes skyrmions \cite{RoldanMolina2015}.
On the other hand, quantum tunneling can open decay channels from a stable to an unstable skyrmion configuration \cite{DerrasChouk2018}.
A purely quantum mechanical treatment could numerically identify quantum skyrmions in systems with Dzyaloshinkii-Moriya interactions (DMI) \cite{Sotnikov2021, Gauyacq2019a} and in frustrated ferromagnets \cite{Lohani2019a} even without chiral interactions.\\
Quantum skyrmions have so far been defined by an adapted classical lattice topological charge \cite{Berg1981} using spin expectation values \cite{Gauyacq2019a}, or by spin triple products  \cite{Sotnikov2021}. Beyond the investigation of ground state properties, tunneling between quantum skyrmion and ferromagnetic states due to the interaction with electrons has been studied \cite{Gauyacq2019a}.
While the transition probabilities between those two configurations are addressed, a reliable procedure for creating quantum skyrmions has not been proposed so far. From spin helices in one-dimensional ferromagnetic chains we know that classical and quantum systems significantly differ in their behavior. While a classical helix can easily be wound up by rotating one edge magnetic moment \cite{Vedmedenko2014}, quantum spin slippage mostly prevents the creation of quantum helices \cite{Posske2019}.
In classical two-dimensional ferromagnetic systems with DMI, rotating the boundary magnetization can create classical magnetic skyrmions \cite{Schaffer2020}. \\
In this work, we explore the creation of quantum skyrmions by manipulating the boundary magnetization of a nanoscale lattice of quantum spins-$1/2$ coupled by ferromagnetic exchange interactions and DMI. First, we classify the ground state by its winding characteristics. Unlike in the classical case, quantum skyrmions can be present at almost vanishing DMI without frustration. 
Those skyrmions, however, have strongly suppressed spin expectation values and are unstable against local perturbations, compared to skyrmion states with larger DMI.
Second, we present an adiabatic \cite{Kato1950} boundary rotation scheme that allows for a controlled creation of metastable quantum skyrmions. We identify viable regimes of creation by varying the DMI and exchange anisotropy. 
Third, we analyze the stability of quantum skyrmions under fluctuations of a local magnetic field and show that pronounced spin expectation values of the individual spins decrease the transition rates between states with different topology.
This manuscript is structured as follows.
In Sec. \ref{SecModel} we introduce the studied quantum spin lattice coupled to a classical boundary. 
In Sec. \ref{SecQSK} we define two quantities to classify the winding properties of the quantum states. Both approach the lattice topological charge in the limit of classical spins, but can capture different properties of the quantum states and can in combination be used to identify stable quantum skyrmions. These quantities are used in Sec. \ref{SecGroundstate} to classify the ground state by its winding characteristics.
In Sec. \ref{SecAdiabaticCreation} we present the adiabatic \cite{Kato1950} boundary rotation scheme that we use to create quantum skyrmions in a controlled manner. We further study which interaction parameter regimes allow for the creation process.
In Sec. \ref{SecStability}, the stability of quantum skyrmion states under local magnetic fluctuation is discussed. Finally, we conclude our results and give an outlook in Sec. \ref{SecConclusion}.

\section{Model}\label{SecModel}
\begin{figure*}[!tb]
	\centering
	\begin{tabular}{c c c c c}
		\begin{picture}(0,0)
		\put(180,75){{\large (a)}}
		\end{picture}
		\begin{picture}(0,0)
		\put(280,75){{\large (b)}}
		\end{picture}
		\begin{picture}(0,0)
		\put(380,75){{\large (c)}}
		\end{picture}
        \includegraphics[width=0.18\textwidth,trim=15 470 10 10,clip]{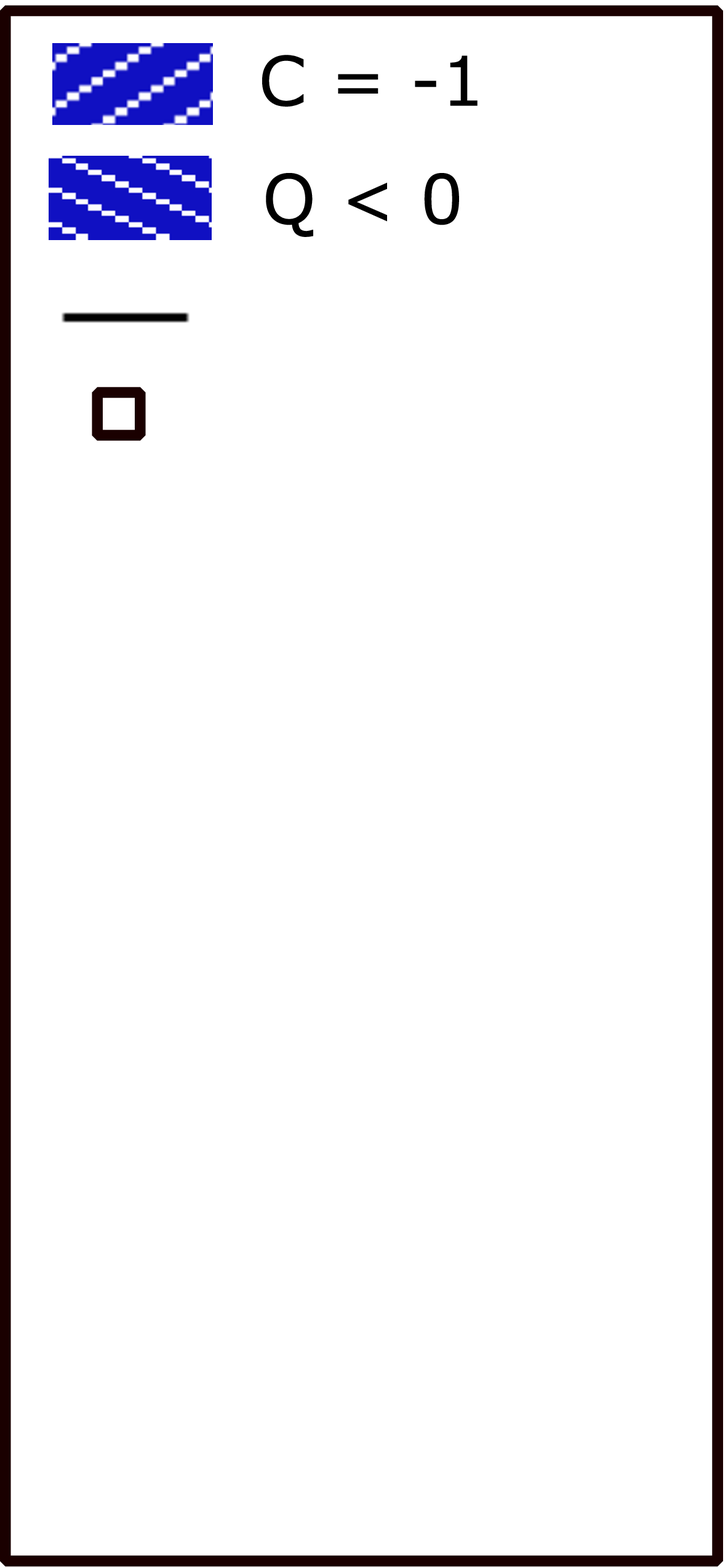}&
        \rule{0pt}{4ex}
        \begin{picture}(0,0)
        \put(-70,29){{\normalsize Gap closure}}
        \end{picture}
        \begin{picture}(0,0)
        \put(-73,16){{\normalsize Classical skyrmion}}
        \end{picture}
		\includegraphics[width=.10\textwidth,trim=0 150 0 0,clip]{WhiteSpace}&
		\includegraphics[width=0.2\textwidth]{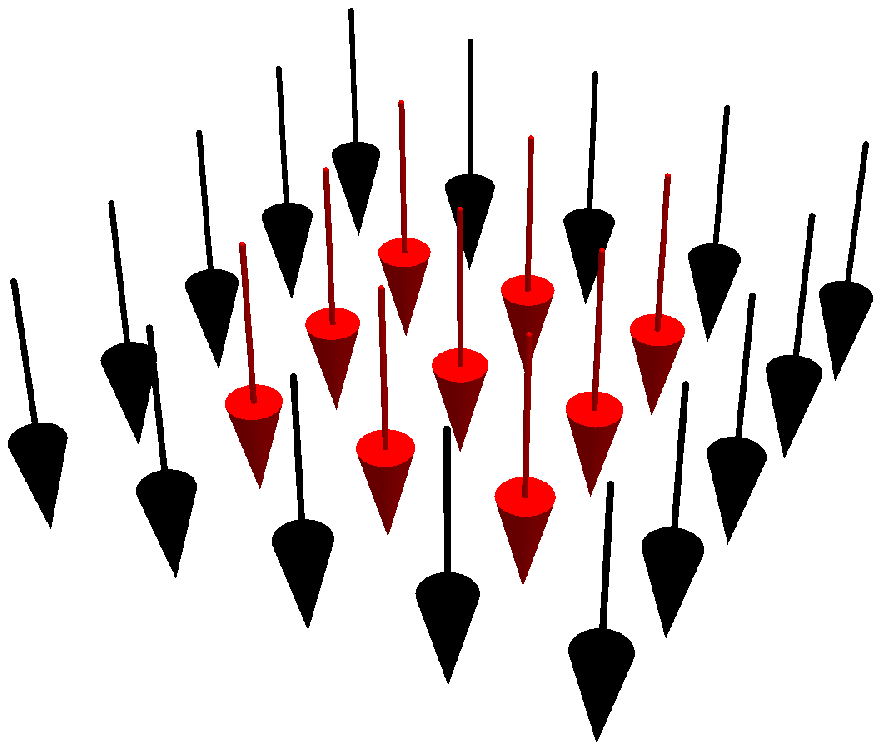}
		& 
		\includegraphics[width=.2\textwidth]{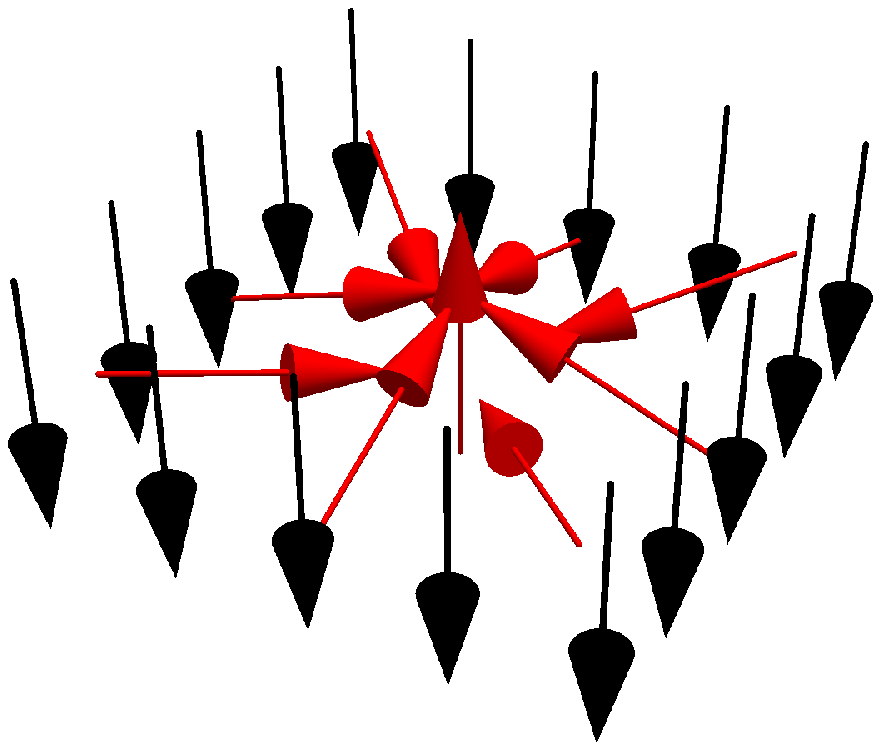}&
		\includegraphics[width=.2\textwidth]{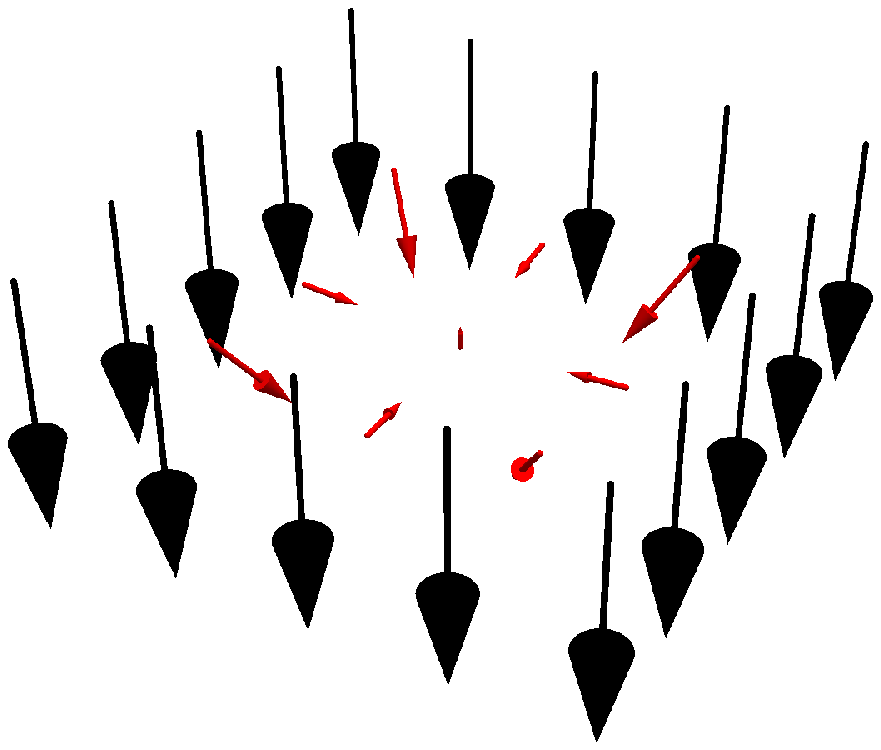}  
	\end{tabular}
	\begin{tabular}{c c}
			\includegraphics[width=.44\textwidth,trim=40 30 0 70,clip]{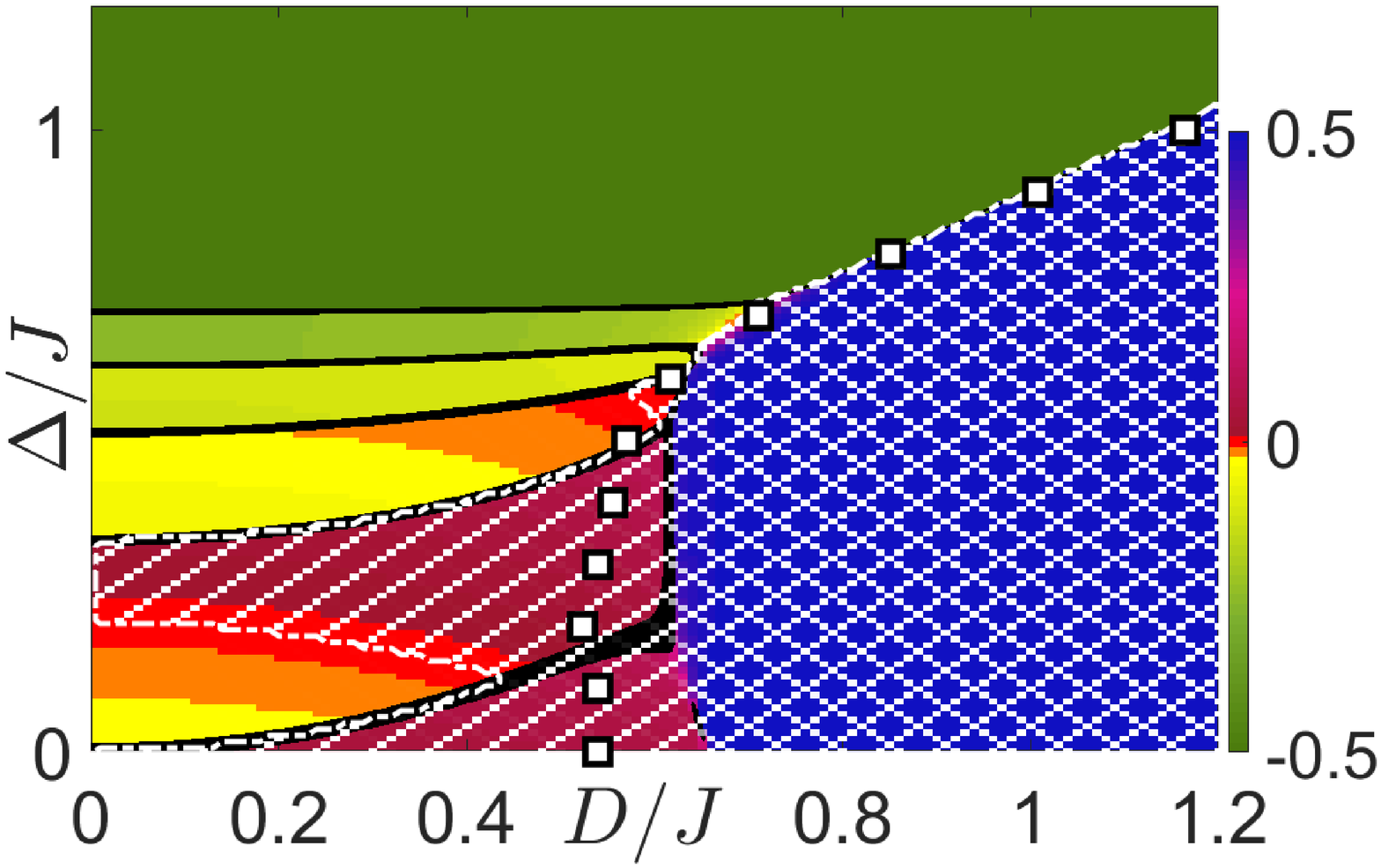}&
			\includegraphics[width=.44\textwidth,trim=40 30 0 70,clip]{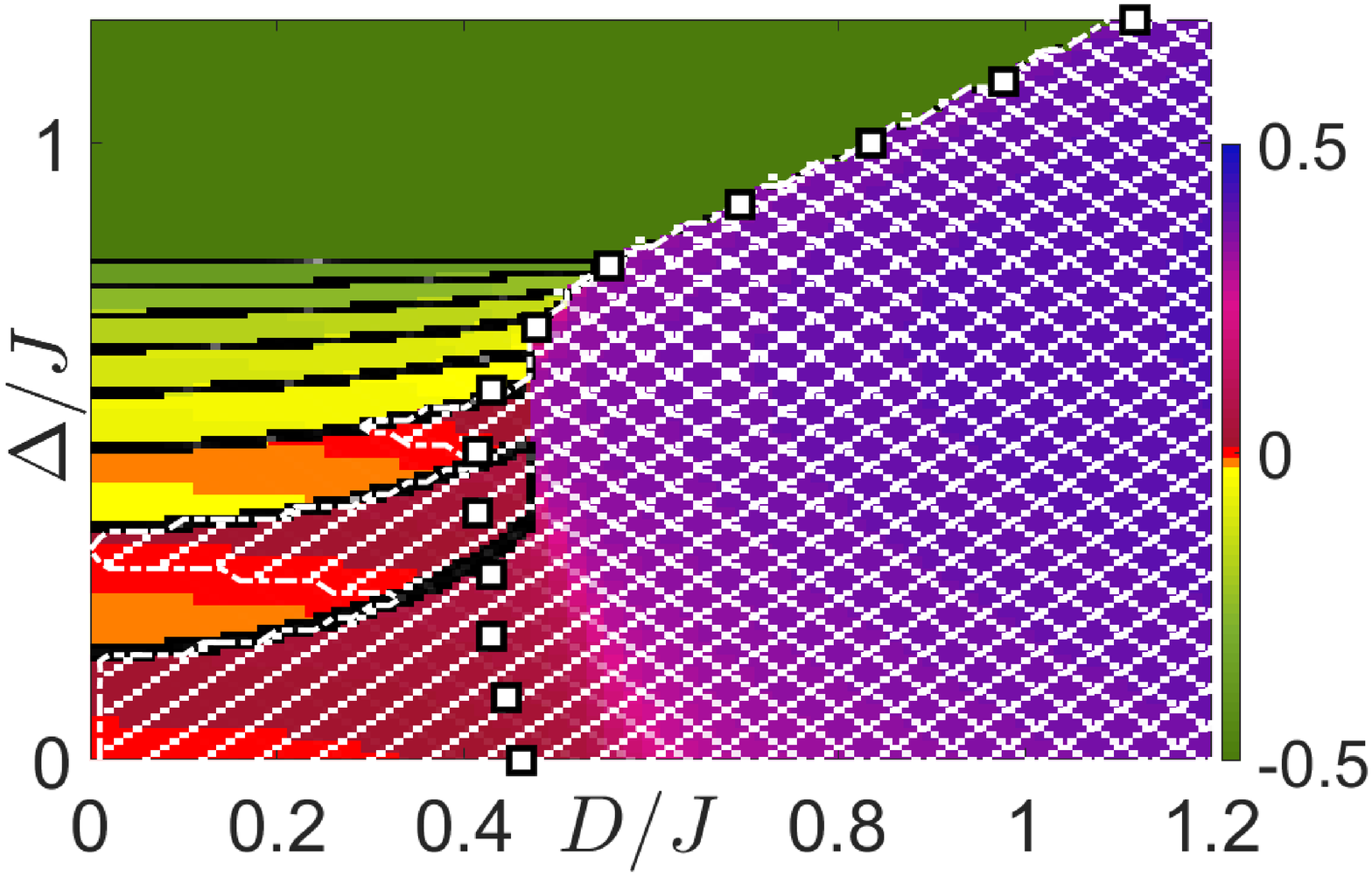} 
			\begin{picture}(0,0)
			\put(-458,130){{\large (d)}}
			\end{picture}
			\begin{picture}(0,0)
			\put(-260,127){{\large $\braket{S^z}_c$}}
			\end{picture}
			\begin{picture}(0,0)
			\put(-234,130){{\large (e)}}
			\end{picture}
			\begin{picture}(0,0)
			\put(-37,127){{\large $\braket{S^z}_c$}}
			\end{picture}
	\end{tabular}
	\caption{Ground states of quantum spin lattices coupled to classical control fields at the boundary.
	(a)-(c) Spin expectation values (red arrows) of a $3\times3$ quantum spin lattice surrounded by a boundary of parallel control fields (black arrows) for representative DMI strengths $D$ and exchange aniostropy $\Delta$: (a) ferromagnet at $D=\Delta=J$, (b) and (c) quantum skyrmions ($C=-1$) at  $D=J,~\Delta=0.5J$ and  $D=0.5J$, $\Delta=0.3J$, respectively.
	(d),(e) Ground state diagram in the $3\times3$ and $4\times4$ quantum spin lattice, including the (average) $z-$component of the spin expectation value of the central spin(s) $\braket{S^z}_c$, the winding parameter $Q$
	and the topological index $C$. The pattern corresponding to $Q$ has an opacity determined by the value of $Q$.
	The value of $Q$ determines the opacity of the corresponding pattern. The black lines mark a degeneracy ($dE<10^{-2}J$) of the ground state.	
	White squares mark the boundary of the regime where classical skyrmions  are the ground state in lattices of magnetic moments of the same system size.  Quantum skyrmions can appear as ground state for infinitesimal DMI in contrast to their classical counterpart
}
	\label{fig:Fig1}
\end{figure*}
We consider an $N\times N$ lattice of interacting quantum spin-$1/2$ coupled to classical magnetic control fields at its boundary. 
The system's Hamiltonian is
\begin{equation}\label{Hamiltonian}
\begin{split}
H=&-J\sum_{<ij>}(S^x_{i}S^x_{j}+S^y_{i}S^y_{j})-\Delta\sum_{<ij>}S^z_{i}S^z_{j}\\&-D\sum_{<ij>}({u}_{ij}\times\hat{z})\cdot(\textbf{S}_i\times \textbf{S}_j),
\end{split}
\end{equation}
with the ferromagnetic Heisenberg exchange coupling $J>0$, the axial  Heisenberg anisotropy $\Delta>0$ and the DMI strength $D$, where $u_{ij}$ is a unit vector pointing from $\textbf{S}_i$ to $\textbf{S}_j$.
In Eq.~(\ref{Hamiltonian}), $\textbf{S}_i=(S^x_i,S^y_i,S^z_i)$ is a vector of spin operators if $i$ indexes a quantum lattice position and a vectorial control field of magnitude $\hbar/2$
if $i$ belongs to the boundary. 
The sum runs over all pairs of nearest neighbors.
A potential realization of the model in a solid state system are nanoskyrmions in a Pd/Fe bilayer islands on Ir(111) decorated at the boundary by ferromagnetic Co/Fe patches \cite{Spethmann2021}, where zero field magnetic skyrmions by boundary tuning were found very recently. Close-by magnetic islands, current-shifted helical magnetic domain walls, or ferroelectric top layers ideally facilitate the desired creation of skyrmions. Furthermore, the model applies to pseudospin lattices, especially in ultracold atoms and noisy intermediate scale quantum computers, where full boundary control is feasible by laser-induced Raman processes, real-time pulse control, and the universal set of gates at the boundary qubits, respectively.

\section{Quantum skyrmions}\label{SecQSK}
The question how to define a skyrmion on a quantum spin lattice has been extensively discussed \cite{Lohani2019a,Gauyacq2019a,Sotnikov2021}.
Realizing the importance of topological arguments in skyrmion science, we employ two quantities with different purposes that both approach the established lattice topological charge \cite{Berg1981} if the quantum spins were replaced by classical magnetic moments, in analogy to Gauyacq et al. \cite{Gauyacq2019a}. These are
\begin{equation}
\begin{rcases}
Q~ \\ 
C~
\end{rcases}
=\frac{1}{2\pi}\sum_{\sigma}\tan^{-1}\left(\frac{\textbf{n}_i(\textbf{n}_j\times\textbf{n}_k)}{1+(\textbf{n}_i\textbf{n}_j+\textbf{n}_i\textbf{n}_k+\textbf{n}_k\textbf{n}_j)}\right),
\end{equation}
where the sum runs over all elementary triangles formed by nearest-neighboring lattice sites $i,~j,~k$.
The winding parameter $Q$ is computed with 
$\textbf{n}_i=2\braket{\textbf{S}_i}/\hbar$,
where $\braket{\textbf{S}_i}=(\braket{S_i^x},\braket{S_i^y},\braket{S_i^z})^T$ is the spin expectation value or the classical magnetic moment. $Q$ is in general a non-integer number which decreases for reduced magnitudes of $\braket{\textbf{S}_i}$.
In contrast, the topological index $C$ relies on the normalized vectors
$\textbf{n}_i=\braket{\textbf{S}_i}/|\braket{\textbf{S}_i}|$. $C$ takes into account only the angular winding properties, irrespective of the magnitude of the spin expectation values and is an integer for parallel boundary fields.
We define each state for which $C=\pm1$ as quantum skyrmion state. The different signs that appear in our results below stem from inverting the polarization of the boundary fields and are unrelated to antiskyrmions.
In addition, the winding parameter $Q$ is an important indicator for the stability of a quantum skyrmion state, as discussed in Sec. \ref{SecStability}.  Quantum skyrmion states with $Q\approx0$ are characterized by almost vanishing spin expectation values and show a very low stability against local perturbations, because minor changes in $\braket{\textbf{S}}$ can result in a spin flip that alters the topological index.
Sotnikov et al. \cite{Sotnikov2021} consider the scalar chirality $\propto\braket{\textbf{S}_i[\textbf{S}_j\times\textbf{S}_k]}$ for identifying quantum skyrmions.  We find that the qualitative features of $Q$ and the scalar chirality coincide, rendering both a good indicator of quantum skyrmions.
Fig.~1~(a)-(c) shows example configurations of spin expectation values of a $3\times3$ quantum spin system which differ in $Q$ and $C$. The first configuration is a ferromagnet with $|\braket{\textbf{S}_i}|\approx\hbar/2$, $Q=0$, $C=0$. 
The second and third configurations are quantum skyrmions ($C=-1$) which differ significantly in $Q$, having $Q=-0.994$ and $Q=-0.034$, respectively. \\
\section{Ground state diagram}\label{SecGroundstate}
We consider the model in Eq.~(\ref{Hamiltonian}) where all boundary fields point downwards.
\begin{figure*}[!tb]
	\centering
	\begin{picture}(0,0)
	\put(0,70){\large (a)}
	\end{picture}
	\begin{tabular}{c c c c c}
		\includegraphics[width=.17\textwidth]{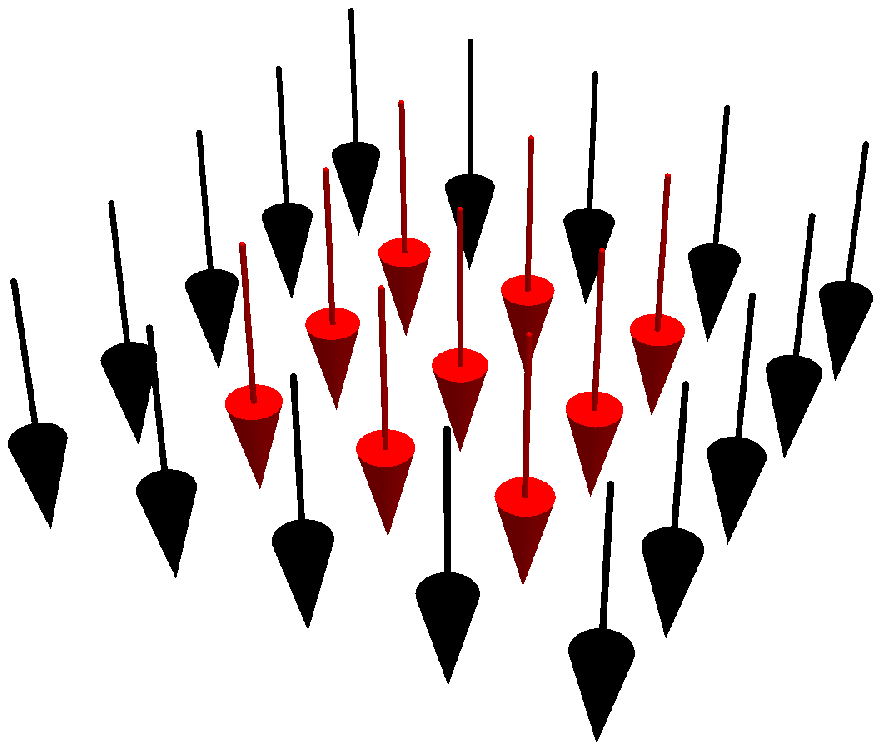}
		\begin{picture}(0,0)
		\put(-85,2){{ $\theta=0$}}
		\end{picture}&
		\includegraphics[width=.17\textwidth]{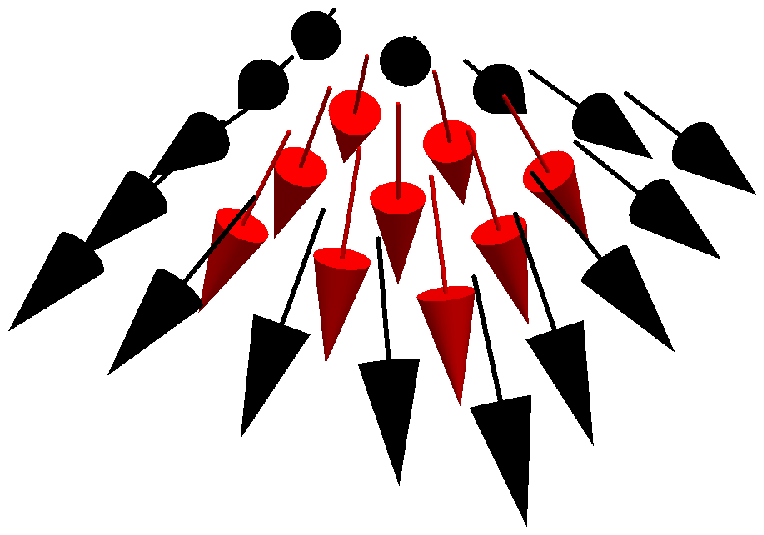}
		\begin{picture}(0,0)
		\put(-85,2){{$\theta=\pi/4$}}
		\end{picture}&
		\includegraphics[width=.17\textwidth]{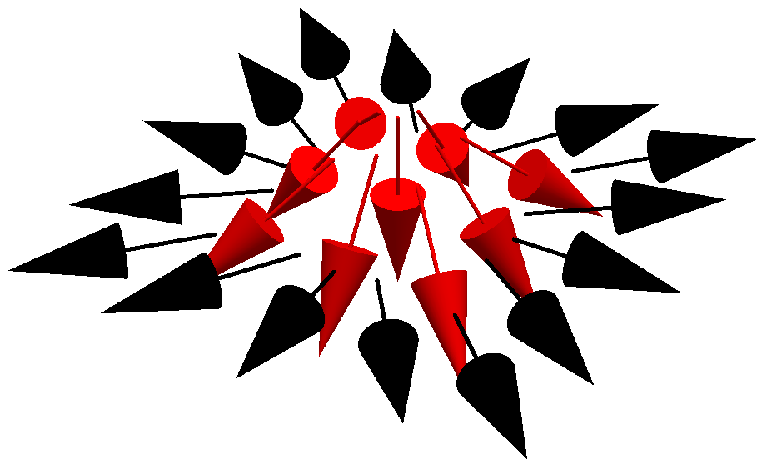}
		\begin{picture}(0,0)
		\put(-85,2){{$\theta=\pi/2$}}
		\end{picture}&
		\includegraphics[width=.17\textwidth]{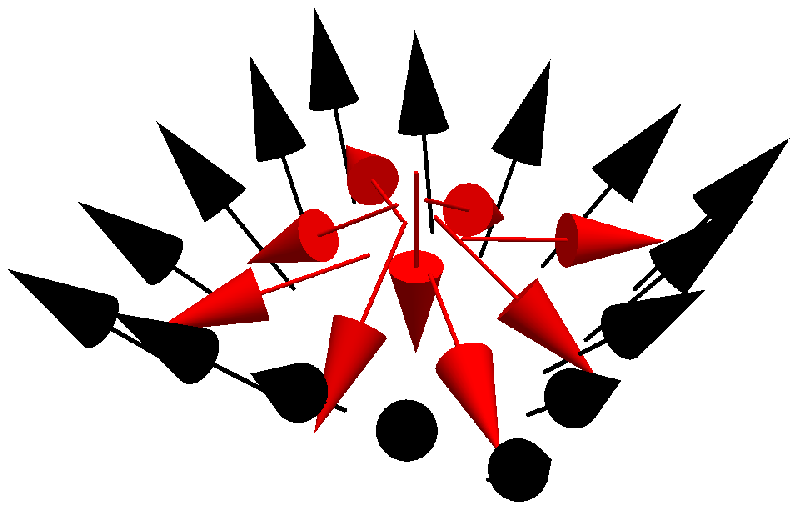}
		\begin{picture}(0,0)
		\put(-85,2){{$\theta=3\pi/4$}}
		\end{picture}&
		\includegraphics[width=.17\textwidth]{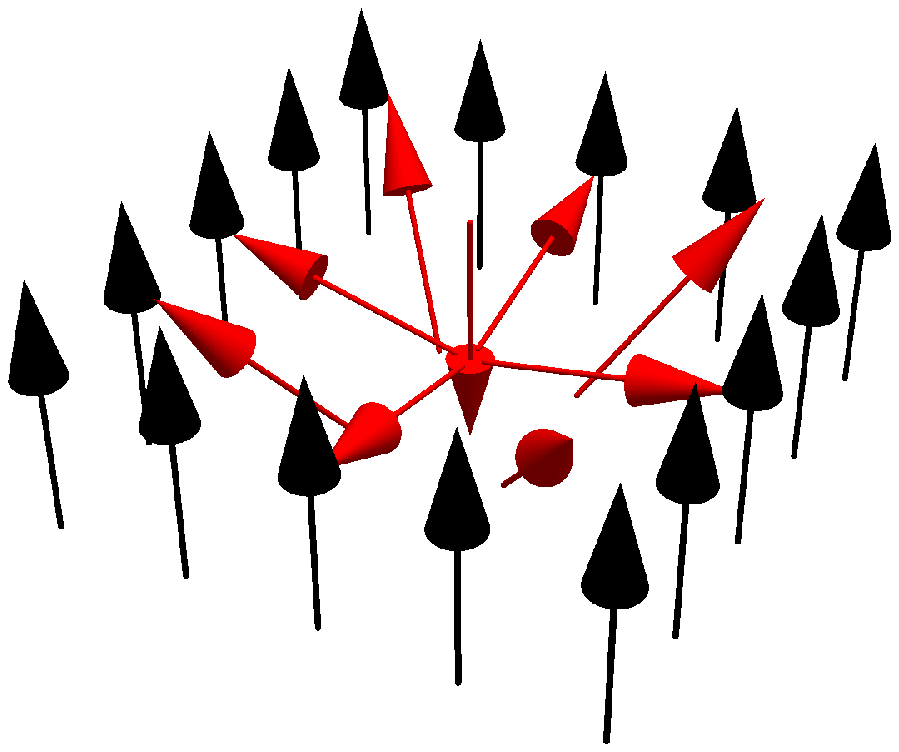}
		\begin{picture}(0,0)
		\put(-85,2){{$\theta=\pi$}}
		\end{picture} \\
		\includegraphics[width=.17\textwidth]{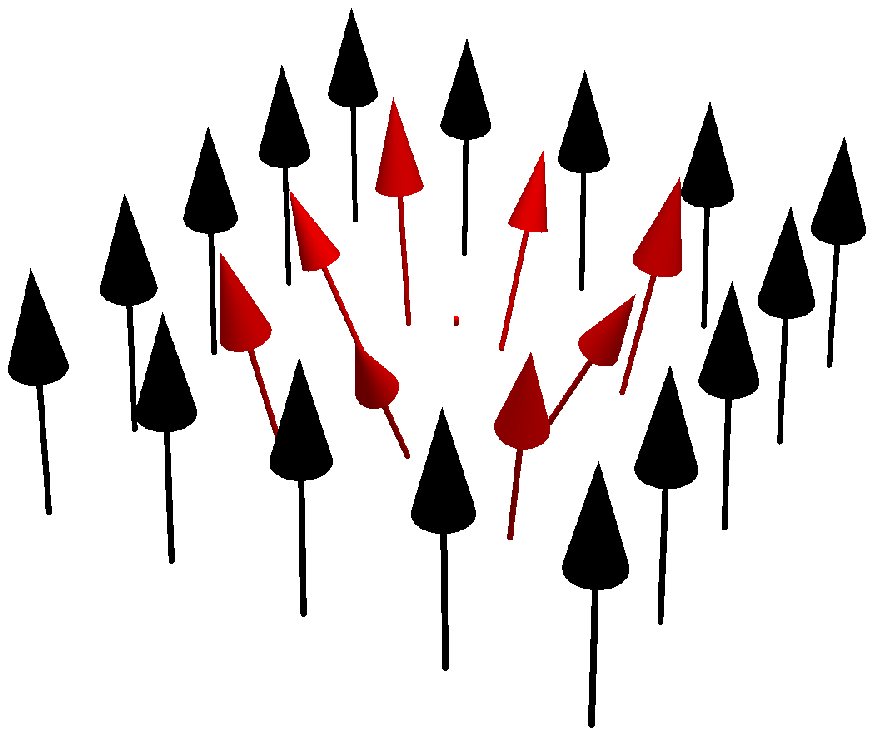}
		\begin{picture}(0,0)
		\put(-85,0){{ $\theta=1.02 \pi$}}
		\end{picture}&
		\includegraphics[width=.17\textwidth]{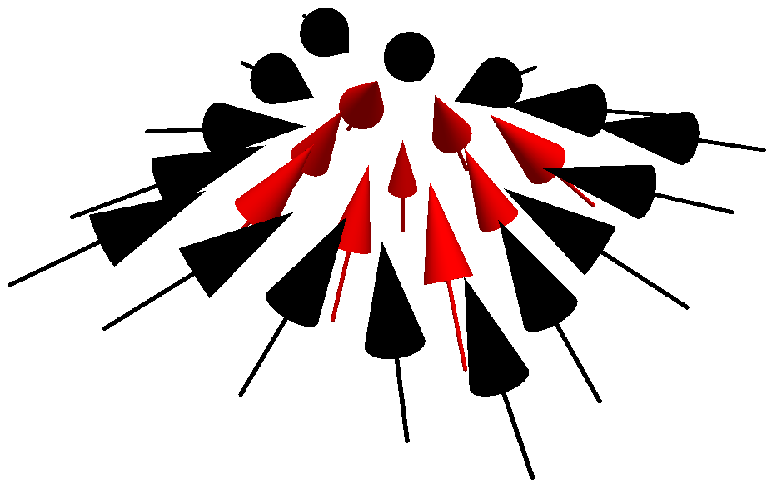}
		\begin{picture}(0,0)
		\put(-85,0){{ $\theta=1.39\pi$}}
		\end{picture}&
		\includegraphics[width=.17\textwidth]{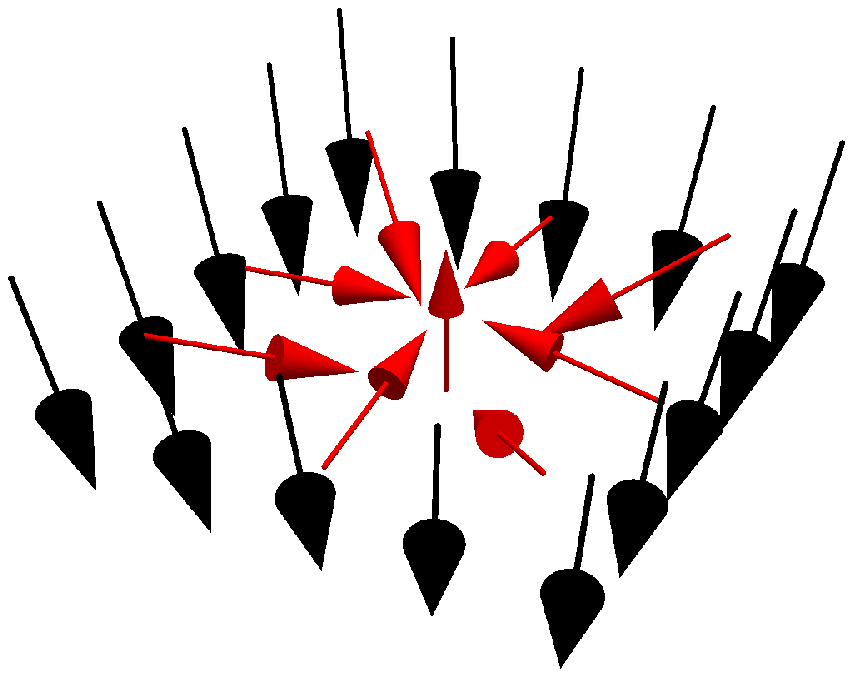}
		\begin{picture}(0,0)
		\put(-85,0){{ $\theta=1.939\pi$}}
		\end{picture}&
		\includegraphics[width=.17\textwidth]{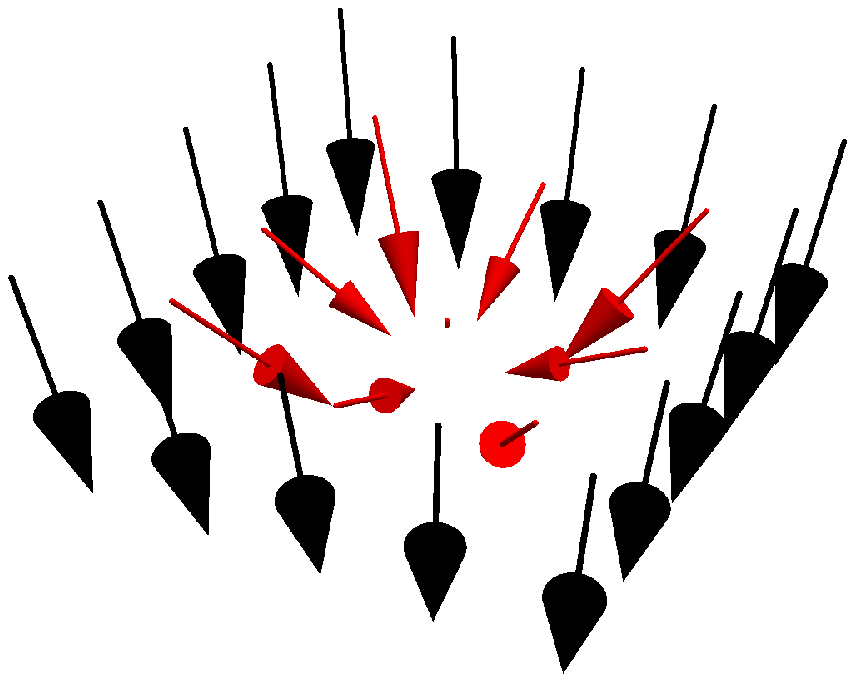}\begin{picture}(0,0)
		\put(-85,0){{ $\theta=1.944\pi$}}
		\end{picture}&
		\includegraphics[width=.17\textwidth]{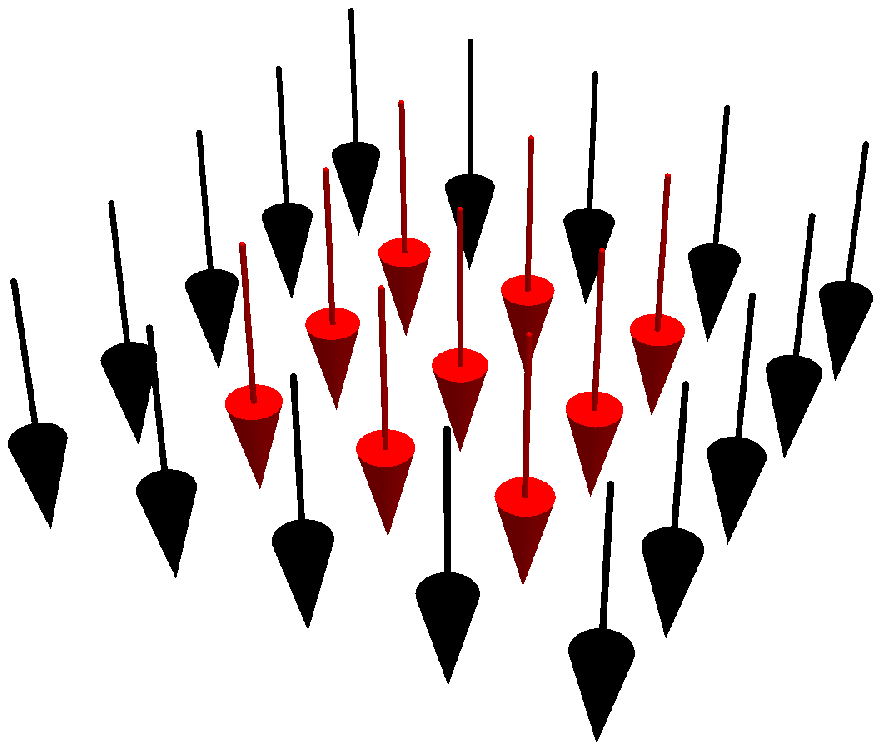}
		\begin{picture}(0,0)
		\put(-85,0){{ $\theta=2\pi$}}
		\end{picture}
	\end{tabular}
	\begin{tabular}{c c c}
		\includegraphics[width=0.33\textwidth,trim=70 0 0 0,clip,keepaspectratio]{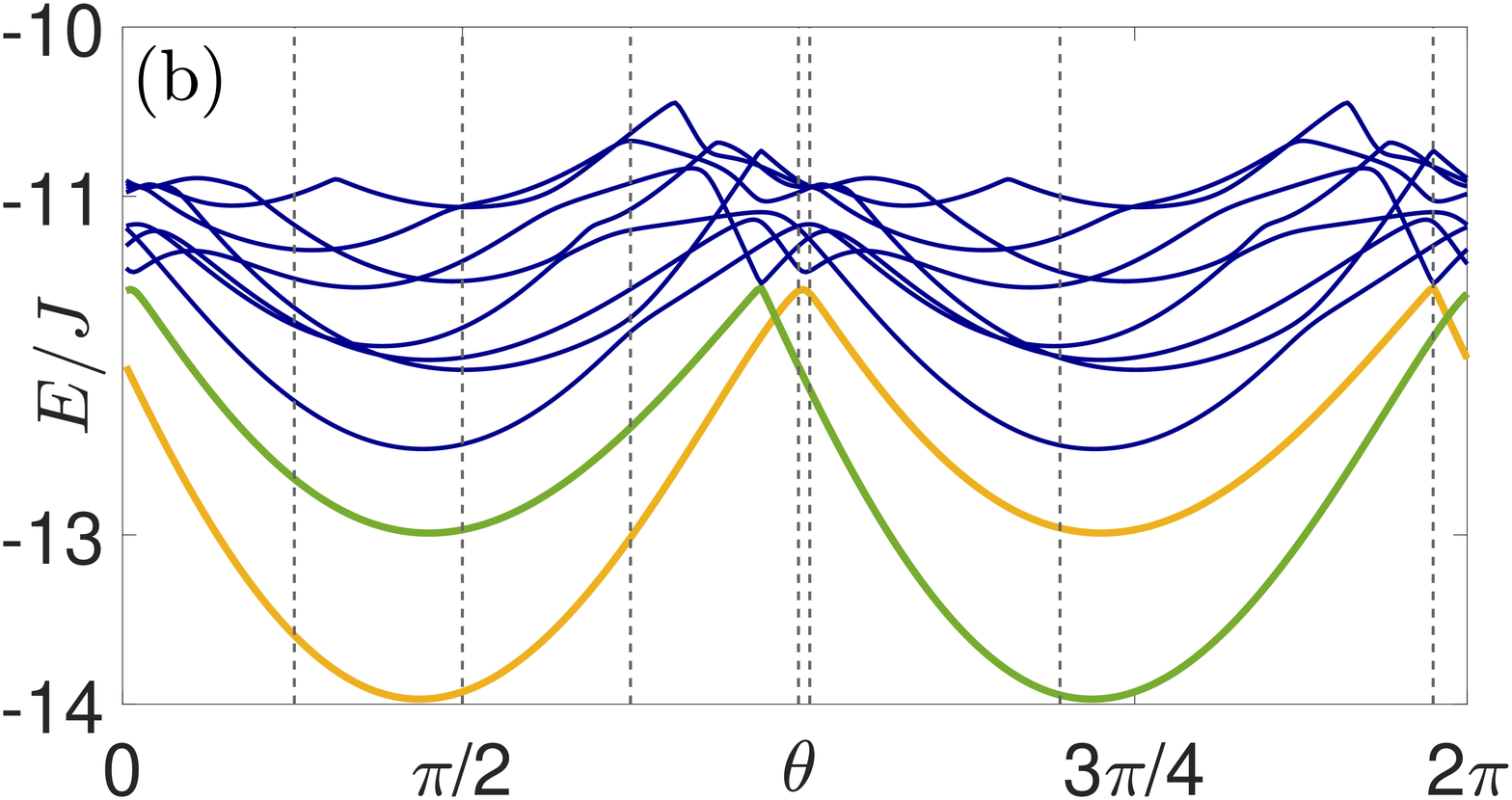}
		\includegraphics[width=0.33\textwidth,trim=70 0 0 0,clip,keepaspectratio]{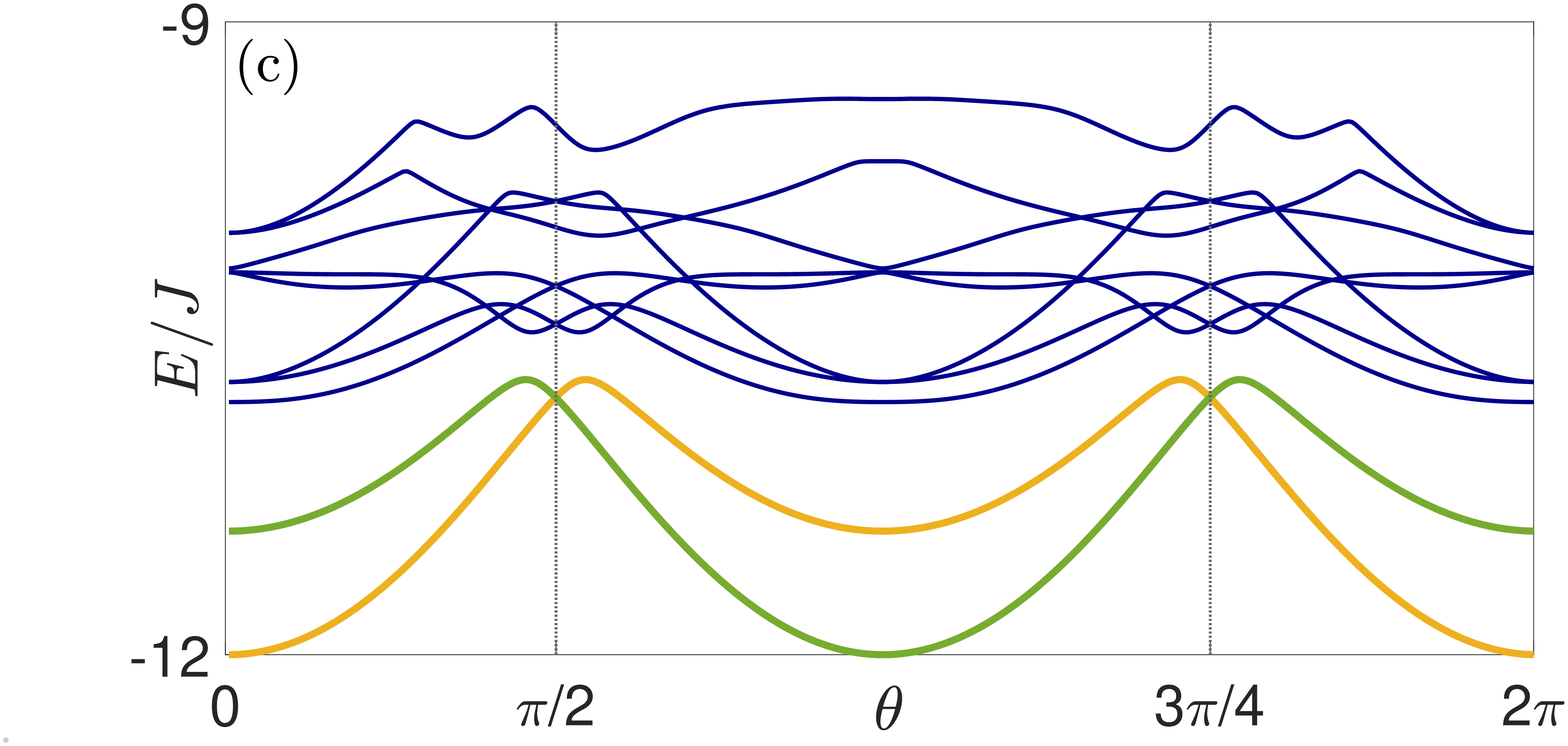}
		\includegraphics[width=0.33\textwidth,trim=70 0 0 0,clip,keepaspectratio]{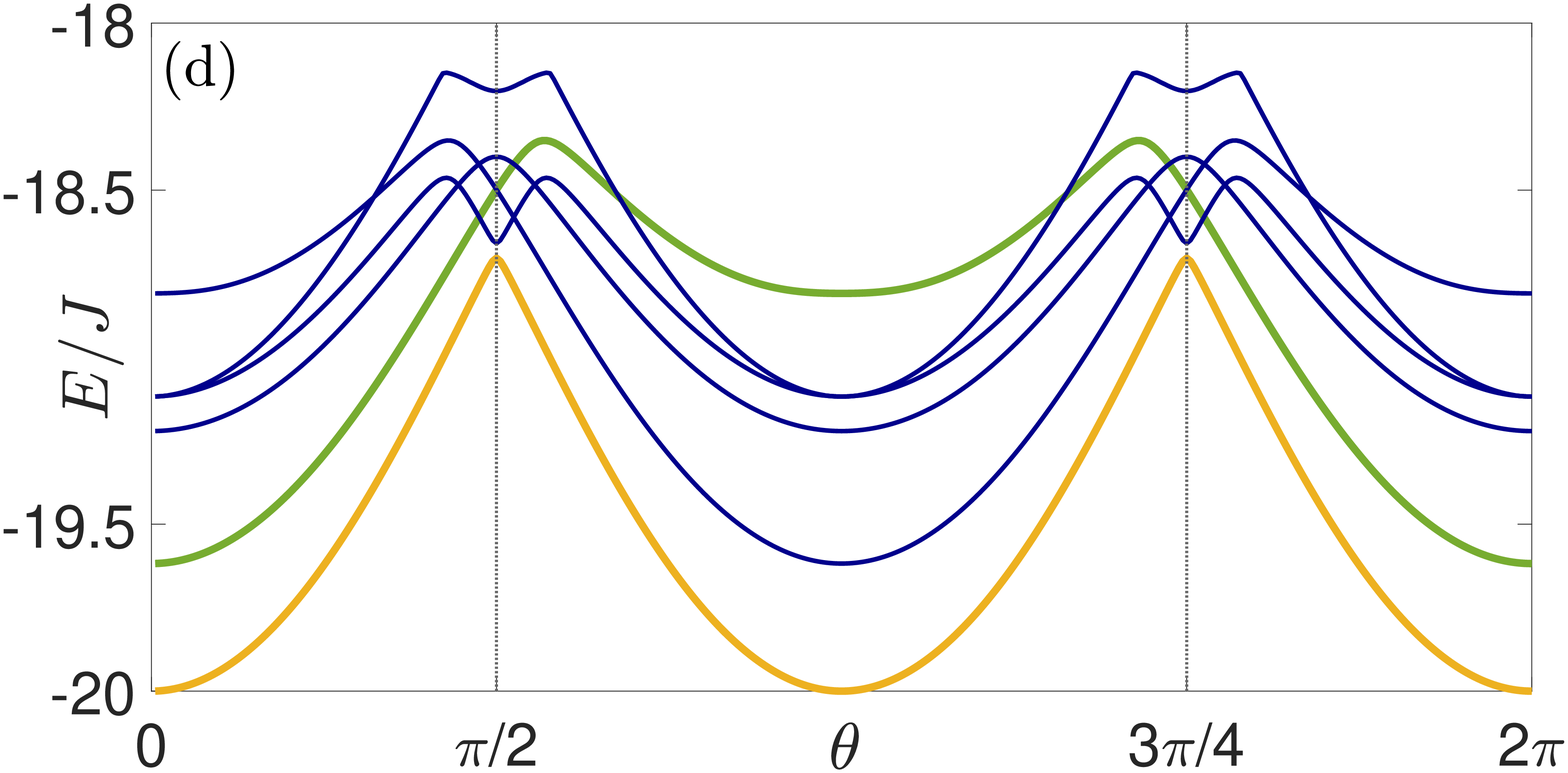}
	\end{tabular}
\caption{Adiabatic evolution of a $3\times3$ quantum spin lattice during the adiabatic rotation of the boundary fields (Eq.~(\ref{AdiabRot})). 
		(a) Magnetic configuration at representative angles $\theta$ for $D=\Delta=J$. A quantum skyrmion is created at $\theta=\pi$ and destroyed at $\theta=1.02\pi$.
		(b) Angle dependence of the energy eigenvalues $E$ and adiabatic evolution for $D=\Delta=J$. Highlighted are the ground state (yellow) and the first excited state (green). Same color denotes adiabatically connected states. Vertical lines mark the angles shown in (a).
		(c) $D=0$ and $\Delta=J$. Vertical lines mark the degeneracy induced by the symmetry $U$ (Eq.~(\ref{USym})).
		(d) $D=0$ and $\Delta=J$ in a $4\times4$ lattice. Here, the symmetry $U$ does not induce 
		a degeneracy of the ground state.}
	\label{2Dquasi}
\end{figure*}
Using exact diagonalization \cite{Kato1950}, we characterize the ground state in dependence on $D$ and $\Delta$ by the corresponding winding parameter $Q$ and topological index $C$. 
We further consider the (average) central $z$-spin expectation value $\braket{S^z}_c=\frac{1}{N_c}\sum_{i\in c}\braket{S^z_i}$, which sums over the $N_c$ central spins, with $N_c=1~(4)$ in the $3\times3$ ($4\times4$) lattice, respectively. For these small lattices, $\braket{S^z}_c$ is a useful indicator for skyrmion structures at $D\neq0$, as $\braket{S^z}_c>0$ corresponds to a quantum skyrmion and  $\braket{S^z}_c<0$ to a trivial state. This yields the ground state diagram shown in Fig.~\ref{fig:Fig1} (d) and (e) for a $3\times3$ and a $4\times4$ quantum spin-$1/2$ lattice, respectively.
The overall influence of the interactions can be explained qualitatively. Larger axial anisotropy $\Delta$ leads to a stronger spin polarization along the $z$-direction due the parallel boundary fields and favors a ferromagnetic spin orientation ($Q=0, C=0$). Larger DMI $D$ favors a noncollinear magnetization at adjacent sites, reducing the energy of a quantum skyrmion state ($Q\approx-1$, $C=-1$).
The regime of small $D$ and $\Delta$ is characterized by strongly decreased spin expectation values such that $Q$ and $C$ can differ significantly. Several energy gap closures indicate energy level crossings in the instantaneous spectrum of $H$ (Eq.~(\ref{Hamiltonian})) when $\Delta$ is changed. Each gap closure gives rise to a changed ground state.
Noteworthy, quantum skyrmions appear even for $D\rightarrow0$, which are, however, characterized by  $Q\approx0$ and the ground state properties are very sensitive to small changes of $D$ and $\Delta$.
In a classical magnetic system with exchange interactions and DMI, isolated skyrmions in confined geometries need a minimal DMI to form a (meta-)stable state, if no other effects such as frustration appear. This is known for skyrmions on a nanodisc \cite{Sampaio2013}. We further used Monte Carlo simulations (see Appendix) to verify that classical skyrmions are only the ground state for $D\geq0.4J$, see white squares in Fig. \ref{fig:Fig1} (b) and (c). \\
For small $D$ and $\Delta$, no energy level crossing is necessary to change the 
topological index $C$ (Fig.~\ref{fig:Fig1}~(d),(e)). 
Under a change of $D$ or $\Delta$, $\braket{S^z}_c$ can shrink to zero and regrow in the opposite direction, which leads to a change of $C$.
This behavior is in contrast to non-interacting gapped electronic systems, where the topological properties can only change by energy gap closures \cite{Ryu2010}, but is due to the fact that the topological number is constructed from the spin expectation values here instead of the Hamiltonian.
In agreement with Ref. \cite{Gauyacq2019a}, we find a shift of the skyrmion ground states to smaller $D$ if the system size is increased.\\
\section{Adiabatic creation of a quantum skyrmions}\label{SecAdiabaticCreation}
In a classical magnetic system, a magnetic skyrmion can be created by a rotation of the magnetization of one of the sample's edge \cite{Schaffer2020}. Here, we explore whether a similar controlled creation of quantum skyrmions is possible.
We first consider the $3\times3$ quantum spin lattice with isotropic exchange ($\Delta=J$) and $D=J$. Initializing the system in the ferromagnetic ground state, the boundary fields are rotated adiabatically with
\begin{equation}\label{AdiabRot}
\textbf{S}_{i}=\left(\frac{\pm\sqrt{w_{i}}}{2}\sin(\theta),\frac{\pm\sqrt{1-w_{i}}}{2}\sin(\theta),\frac{-1}{2}\cos(\theta)\right),
\end{equation}
where $0<w_{i}<1$ are position-dependent weights and $i\in\text{boundary}$. Various choices of the $w_i$ can lead to a successful creation of a quantum skyrmion (see Appendix). For concreteness, we focus on a symmetric rotation of all edges shown in Fig \ref{2Dquasi} (a) with values of the $w_i$ given in the Appendix.
This rotation creates a quantum skyrmion with $Q=0.799$ at $\theta=\pi$ (Fig.~\ref{2Dquasi}~(a)). The process is also shown the Supplemental Video. 
The corresponding evolution of the instantaneous energy spectrum is depicted in Fig.~\ref{2Dquasi}~(b). 
The adiabatically evolved ground state crosses the first excited state taking the second position in the energy spectrum at $\theta=\pi$, thereby transforming into a quantum skyrmion ($C=1$). During the first half of the rotation, the spin expectation values stay close to their maximal magnitude of $\hbar/2$. Upon continuing the rotation, the central spin expectation value vanishes, which leads to a change from a skyrmion-like to a trivial configuration ($C\approx0$, $C$ not quantized due to non-parallel boundary fields).  
At $\theta=2\pi$, the ground state is back in the ferromagnetic initial configuration.
Notice that an exact level crossing here amounts to  changing eigenvectors, in contrast to a usual Landau-Zener-Stückelberg transition, where the case of exact crossing corresponds to unchanged eigenvectors.
\\
A different choice of the interaction parameter $D$ influences both the initial configuration of the quantum states as well as the system's evolution under the rotation.
For isotropic exchange ($\Delta=J$), the crossing between the ground and the first excited state persists for all considered values of $D$. However, a quantum skyrmion is only created in the regime  $0.94J\leq D \leq 1.14J $, where the ground state is initially trivial and the first excited state is a skyrmion. 
The energy evolution for $D=0$ is depicted in Fig.~\ref{2Dquasi}~(c).
For $D=0$, an energy level crossing is caused by the unitary symmetry 
\begin{equation}\label{USym}
U = P_{x\Leftrightarrow y}U_{rot},
\end{equation} 
which commutes with $H$ for $\theta=\pi/2$ and $\theta=3\pi/4$.
Here,  $U_{rot}=\exp(i\pi\frac{1}{\sqrt{2}}\sum_{<i,j>}(S_{i}^x+S_{j}^y))$ is a spin rotation by an angle $\pi$ around the axis $\frac{1}{\sqrt{2}}(1,1,0)^T$ and $P_{x\Leftrightarrow y}$ is a permutation matrix which exchanges the quantum spins at positions $i=(x,y)$ and $j=(y,x)$.
This symmetry is not present for $D\neq0$. However, the crossing persists for all values of $D$ in extensive numerical search, only being shifted to larger twisting angles $\theta$ for increasing $D$. Hence, the degeneracy of the ground state at some twisting angle is a generic property for isotropic exchange despite the lack of symmetry protection.\\
\begin{figure}[!tb]
\centering
	\includegraphics[width=0.5\textwidth,trim=0 20 0 0,clip,keepaspectratio]{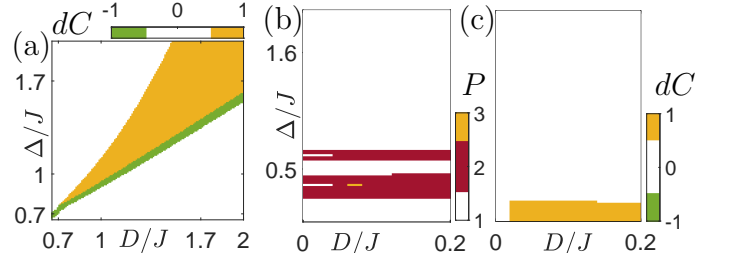}
	\begin{picture}(0,0)
		\put(-116,85){\large (a)}
	\end{picture}
	\begin{picture}(0,0)
		\put(-24,95){\large (b)}
	\end{picture}
	\begin{picture}(0,0)
		\put(47,95){\large (c)}
	\end{picture}
	\begin{picture}(0,0)
	\put(-110,93){\large $dC$}
	\end{picture}
	\begin{picture}(0,0)
	\put(42,70){\large $P$}
	\end{picture}
	\begin{picture}(0,0)
	\put(112,70){\large $dC$}
	\end{picture}
	\caption{(a) Regime of creation and destruction of quantum skyrmions. Depicted is the difference $dC$ in the absolute value of the topological index of the adiabatically evolved ground state for the $3\times3$ quantum spin lattice. $dC=1$ corresponds to the regime of quantum skyrmion creation at a twisting angle $\theta=\pi$, $dC=-1$ to a regime of quantum skyrmion destruction. The ground state always reaches the first excited state in the parameter regime shown in (a).
	(b)	Position $P$, corresponding to the $P$th excited state that the adiabtically evolved ground state reaches in the instantaneous spectrum at $\theta=\pi$. 
	(c) Skyrmion creation for low $D$.}
	\label{fig:Fig4}
	\end{figure}
Quantum skyrmions can also be created for anisotropic exchange ($ \Delta\neq0$). For $\Delta$ close to $J$ or large $D$, the system's behavior under a rotation stays the same, leading to the creation of a quantum skyrmion if the first excited state is initially skyrmionic. Fig.~\ref{fig:Fig4}~(a) depicts the change of the topological index $dC = |C(\theta=\pi)| - |C(\theta=0)|$, defining regimes of quantum skyrmion creation and destruction.
Further, states beyond the first excited state can be reached by the adiabatic evolution of the ground state.
For $\Delta<0.7J$, Fig.~\ref{fig:Fig4}~(b) shows that states up to the third excited state are reached at the twisting angle $\theta=\pi$.
However, these states are not quantum skyrmions, see Fig.~\ref{fig:Fig4}~(c).
We further find, that for $\Delta>2$, the avoided crossings with energetically higher states decrease in size, requesting decreasing step size in the simulation.
 \begin{figure*}
	\begin{minipage}{0.18\textwidth}
		\centering
		\begin{picture}(0,0)
		\put(2,65){{\large (a)}}
		\end{picture}
		\vspace{-0.5cm}
		\begin{tabular}{|c|c|c|}
			\hline
			\textbf{index} & \textbf{Q} & \textbf{C} \\
			\hline
			
			1 & 0 & 0 \\
			
			2 & -0.001 & 0 \\
			
			3 & -0.003 & 0 \\
			
			\arrayrulecolor{red}\hline
			4 & -0.77 & -1 \\
			\hline
			\arrayrulecolor{black}
			
			5 & -0.023 & 0 \\
			\arrayrulecolor{orange}\hline
			6 & -0.19 & -1 \\
			\hline
			\arrayrulecolor{black}
			7 & -0.02 & 0 \\
			
			8 & -0.04 & 0 \\
			
			9 & -0.172 & -1 \\
			
			\hline
		\end{tabular}
	\end{minipage}
	\begin{minipage}{0.23\textwidth}
		
		\includegraphics[width=\linewidth,
		height=\textheight,keepaspectratio]{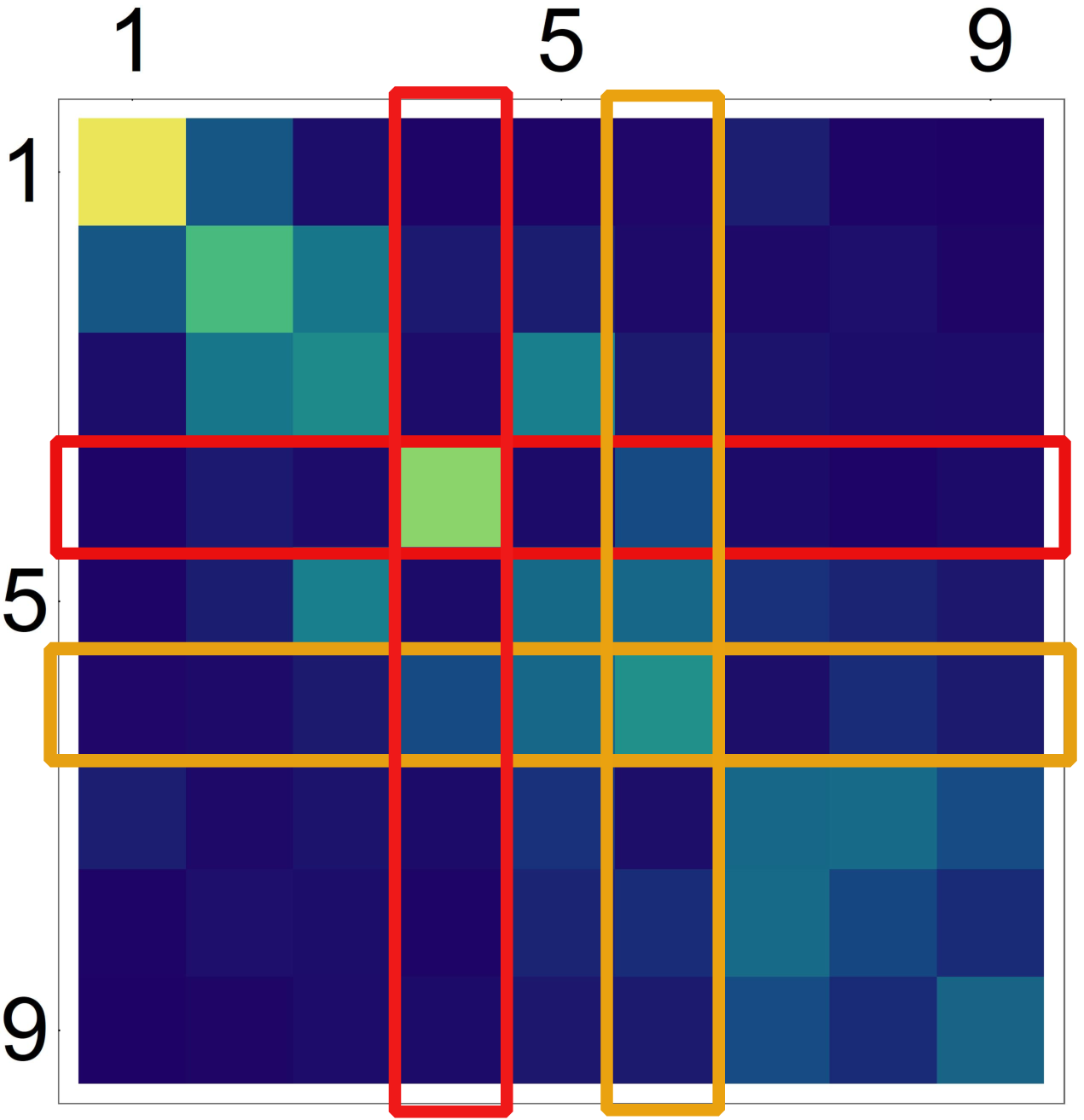}
	\end{minipage}
	\begin{minipage}{0.05\textwidth}
		\begin{picture}(0,0)
		\put(-20,0){{\large $\frac{\hbar\Gamma_{j,k}}{2\pi|\alpha \tilde{B}_{\text{max}}|^2}$}}
		\end{picture}
		
		\vspace*{3 ex}
		\includegraphics[width=\linewidth,
		height=0.17\textheight,keepaspectratio]{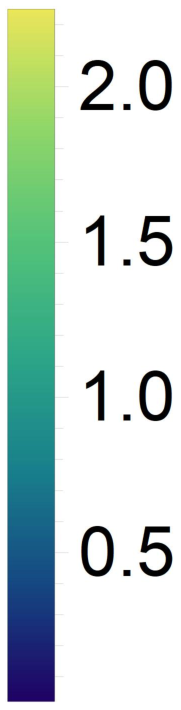}
	\end{minipage}
	\hspace{0.6cm}
	\vline
	\hspace{0.0cm}
	\begin{minipage}{0.18\textwidth}
		\centering
		\begin{picture}(0,0)
		\put(2,65){{\large (b)}}
		\end{picture}
		\vspace{-0.5cm}
		\begin{tabular}{|c|c|c|}
			\hline
			\textbf{index} & \textbf{Q} & \textbf{C} \\
			\hline
			
			1 & -0.0018 & 0 \\
			
			2 & -0.0003 & 0 \\
			\arrayrulecolor{NewOrange}\hline
			3 & -0.010 & -1 \\
			\hline\arrayrulecolor{black}
			4 & -0.00008 & 0 \\
			
			5 & -0.0043 & -1 \\
			\arrayrulecolor{NewPink}\hline
			6 & 0 & 0 \\
			\hline\arrayrulecolor{black}
			7 & -0.0066 & 0 \\
			
			8 & -0.0001 & 0 \\
			
			9 & -0.0029 & 0 \\
			
			\hline
		\end{tabular}
	\end{minipage}
	\begin{minipage}{0.23\textwidth}
		
		\includegraphics[width=\linewidth,
		height=\textheight,keepaspectratio]{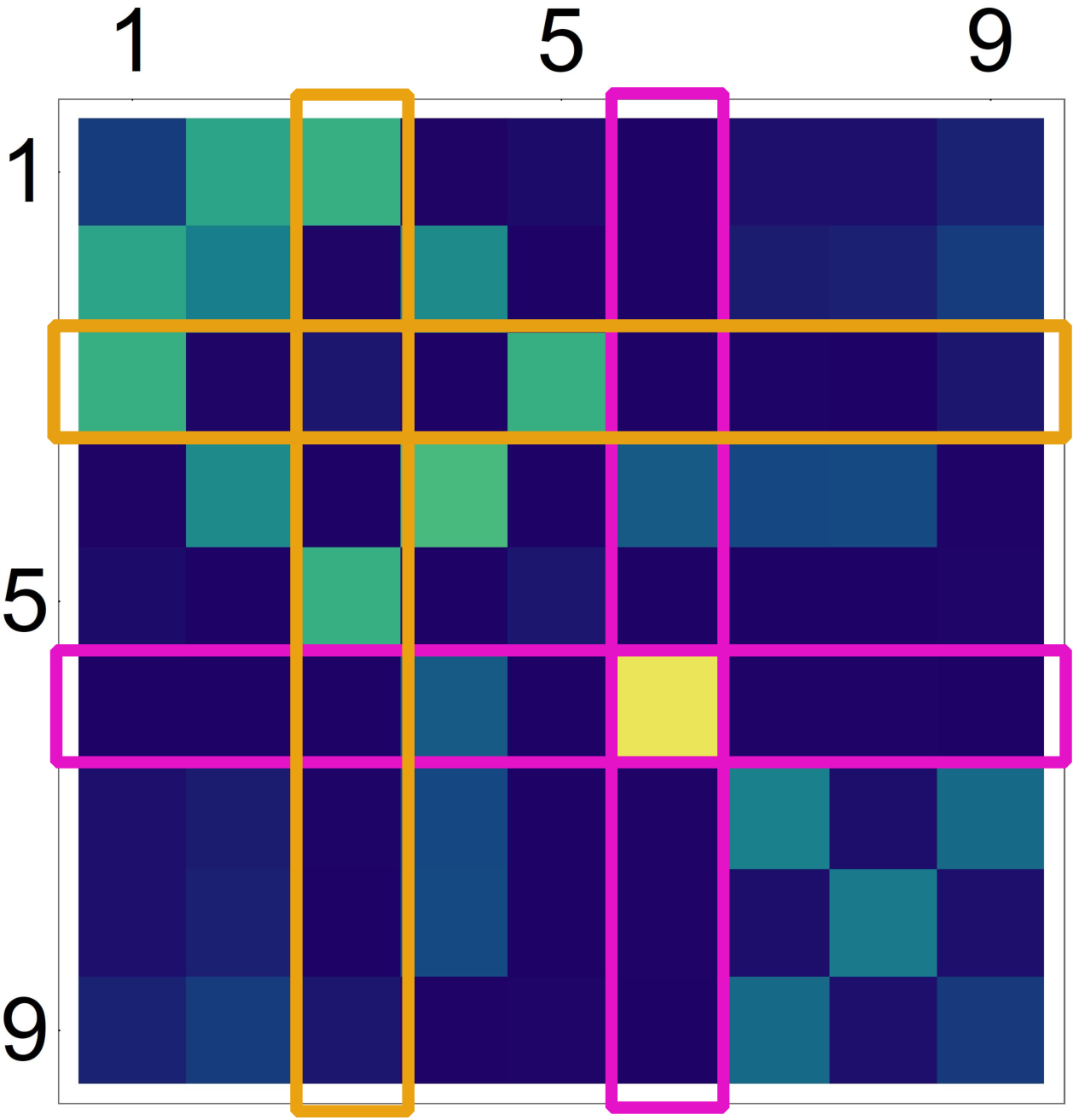}
	\end{minipage}
	\begin{minipage}{0.05\textwidth}
		\begin{picture}(0,0)
		\put(-20,0){{\large $\frac{\hbar\Gamma_{j,k}}{2\pi|\alpha \tilde{B}_{\text{max}}|^2}$}}
		\end{picture}
		
		\vspace*{3 ex}
		\includegraphics[width=\linewidth,
		height=0.17\textheight,keepaspectratio]{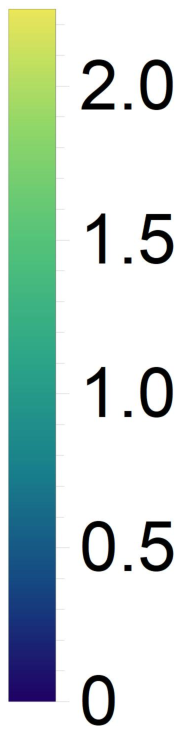}
	\end{minipage}
	\caption{Upper bound of transition rates
		$\Gamma_{j,k}$ (Eq.~(\ref{Transi})) between two eigenstates
		$\ket{\Psi_j}$ and $\ket{\Psi_k}$ of the unperturbed Hamiltonian and the
		$Q-$ and $C-$ values for those states in a $3\times3$ quantum spin lattice
		affected by independent magnetic fluctuation at each site for (a) $D=0.7J$,
		$\Delta=0.8J$ and (b) $D=0.2J$, $\Delta=0.5J$, respectively.
		Quantum skyrmions with a large magnitude of $Q$ and ferromagnetic
		states with $\braket{\textbf{S}}\approx\frac{\hbar}{2}$ have suppressed
		transition matrix elements.
	}
	\label{Sup:Trans1}
\end{figure*}
For $D=0$, the crossings with energetically higher states are still explained by the symmetry $U$ (Eq.~(\ref{USym})). For anisotropic exchange far from $\Delta\approx J$, we find that the number of crossings is not constant but varies with $D$. \\
Not all rotation schemes result in a successful quantum skyrmion creation, see Appendix. 
In contrast to classical systems \cite{Schaffer2020}, it is in general not sufficient to rotate the magnetization of one edge of the quantum spin lattice, as then, required energy level crossings appear only at isolated interaction parameters, see Appendix. \\
We further explore the possibility to create quantum skyrmions in the $4\times4$ lattice, considering a symmetric edge rotation (Eq.~\ref{AdiabRot}). Compared to the $3\times3$ lattice, the system's behavior is rather different. For $D=0$, the Hamiltonian obeys the same symmetry $U$ (Eq.~(\ref{USym})) at $\theta=\pi/2$ and $\theta=3\pi/2$. This does, however, not lead to a degeneracy of the ground state (see Fig.~\ref{2Dquasi}~(d)) but leaves it invariant. For isotropic exchange, this absence of level crossings is numerically generic (see Appendix), preventing the creation of quantum skyrmions. 
This generality is lost for anisotropic exchange,
where we find exemplary gap closures for specific $\Delta$ and $D$. 
\section{Stability of quantum skyrmions}\label{SecStability}
One of the reasons why classical magnetic skyrmions are promising
candidates for technological applications is their robustness against local
perturbations (see Refs. \cite{Nagaosa2013, Vedmedenko2019}).
Here, we study the stability of quantum skyrmions against small time-fluctuating magnetic fields $\vec{B}_{i}(t)$ that couple locally to each spin $\textbf{S}_i$ of the system. This is described by the perturbation Hamiltonian
$H_{pert, i}^{\lambda}=\alpha B_i^{\lambda}(t) S_i^{\lambda}$,
with the coupling constant $\alpha$. 
To compute an upper bound of the transitions matrix elements between the two considered quantum states $\ket{\Psi_j}$ and $\ket{\Psi_k}$, we make the following two approximations. 
First, we assume that the local magnetic fields fluctuations are spatially uncorrelated. According to Fermi's golden rule, the transition rates induced by a single of these fields then is $\frac{2\pi |\alpha \tilde{B}_i^\lambda(j,k) |^2}{\hbar}|\langle \Psi_j| S_i^\lambda | \Psi_k\rangle|^2$, where $\tilde{B}_i(j,k)$ is the Fourier component of the time-dependent magnetic field at the frequency that corresponds to the energy difference of the eigenstates $|\Psi_j\rangle$ and $|\Psi_k\rangle$. Second, one $\tilde{B}^\lambda_i(j,k)$ is maximal, i.e., $\tilde{B}_{\text{max}}=\max_{i,j,k,\lambda}{\left(\tilde{B}^\lambda_i(j,k)\right)}$, which take to calculate an upper bound for the transition rates 
\begin{equation}
\gamma_{j,k}^{S_i^{\lambda}} = \frac{2\pi|\alpha
		\tilde{B}_{\text{max}}|^2}{\hbar}|\bra{\Psi_j} S_{i}^{\lambda}\ket{\Psi_k}|^2.
\end{equation}
Consequently, an upper bound of the
total transition rate is computed by summing up the individual
contributions according to
\begin{equation}\label{Transi}
\Gamma_{j,k}=\sum_{\lambda,i}\gamma_{j,k}^{S_i^{\lambda}}.
\end{equation}
In Fig.~\ref{Sup:Trans1}~(a), we depict the upper bounds of the transition rates
$\Gamma_{j,k}$ between the nine energetically lowest states of a
$3\times3$ quantum spin lattice with $D=0.7J$ and $\Delta=0.8J$, together with the winding parameter $Q$ and the topological index
$C$.
The data include quantum skyrmion states ($C=-1$), which differ
significantly in their $Q$-value.
In comparison to other states, the 3rd excited state with $Q=-0.77$ has strongly suppressed
transition rates. The only significant
transition takes place into the 5th excited state, which is a
quantum skyrmion state with non-vanishing $Q$ as well. In contrast, the 5th and
the 8th excited state, which are as well quantum skyrmions but with small $Q$,
exhibit significant transition rates into trivial states ($C=0$).
The reduced stability of quantum skyrmions with $|Q|\ll 1$ is also
apparent in Fig.~\ref{Sup:Trans1}~(b) for a $3\times3$ lattice with
$D=0.2J$ and $\Delta=0.5J$. There, the second excited state, which is a quantum
skyrmion with $Q=-0.01$, has frequent transitions into the trivial ground
state.
Not only quantum skyrmion states, but also ferromagnetic states with large
$\braket{\textbf{S}}$ show an increased stability as, e.g., the 5th
excited state in Fig.~\ref{Sup:Trans1}~(b).\\
In conclusion, we see that both quantum skyrmion and trivial states with
pronounced spin expectation values
$\braket{\textbf{S}}\approx\hbar/2$ show decreased transition rates,
signifying a robustness against external magnetic fluctuations. However,
they still show non-vanishing transitions into states with the same $C$.
Thus, a step-wise decay process could take place.
\section{Conclusion}\label{SecConclusion}
We explore the controlled creation of nanoscale magnetic skyrmions in a quadratic lattice of quantum spins coupled by DMI and exchange interactions manipulated by control fields the boundary. In contrast to classical systems without frustrated magnetism,  a quantum skyrmion can be the ground state even for infinitesimal DMI. The appearing quantum skyrmions are characterized by significant differences in the magnitudes of their spin expectation values and show large differences in stability.
Quantum skyrmions can further be created as metastable excitations from a ferromagnetic ground state by a tailored adiabatic rotation of the boundary fields. 
For isotropic exchange, necessary energy level crossings are numerically generic in the $3\times3$ lattice and protected for $D=0$ by a unitary symmetry. Instead, in the $4\times4$ lattice with isotropic exchange, we find a generic absence of level crossings, preventing the creation of topologically nontrivial spin structures. 
It would be interesting, even though numerically challenging, to generalize the findings of the $3\times3$ and $4\times4$ lattices to larger lattices.
We show that transition between quantum states of different $C$ with pronounced spin expectation values is significantly reduced.
The controlled creation of quantum skyrmions is a prerequisite for quantum skyrmion applications. For classical skyrmion memory devices it is necessary to know whether the stability and control of skyrmions persists when entering the quantum regime.
Also, beyond classical skyrmion applications the proposed concept allows for a controlled switch between two quantum states with different topological properties.\\
\section{Acknowledgments}
PS and MT acknowledge funding by the Cluster of Excellence 
‘Advanced Imaging of Matter’ (EXC 2056—project ID 390715994) of the Deutsche 
Forschungsgemeinschaft (DFG). TP acknowledges funding by the DFG (project no. 420120155). MS acknowledges funding by the DFG (project no. 403505707).
\FloatBarrier

\begin{thebibliography}{34}%
\makeatletter
\providecommand \@ifxundefined [1]{%
 \@ifx{#1\undefined}
}%
\providecommand \@ifnum [1]{%
 \ifnum #1\expandafter \@firstoftwo
 \else \expandafter \@secondoftwo
 \fi
}%
\providecommand \@ifx [1]{%
 \ifx #1\expandafter \@firstoftwo
 \else \expandafter \@secondoftwo
 \fi
}%
\providecommand \natexlab [1]{#1}%
\providecommand \enquote  [1]{``#1''}%
\providecommand \bibnamefont  [1]{#1}%
\providecommand \bibfnamefont [1]{#1}%
\providecommand \citenamefont [1]{#1}%
\providecommand \href@noop [0]{\@secondoftwo}%
\providecommand \href [0]{\begingroup \@sanitize@url \@href}%
\providecommand \@href[1]{\@@startlink{#1}\@@href}%
\providecommand \@@href[1]{\endgroup#1\@@endlink}%
\providecommand \@sanitize@url [0]{\catcode `\\12\catcode `\$12\catcode
  `\&12\catcode `\#12\catcode `\^12\catcode `\_12\catcode `\%12\relax}%
\providecommand \@@startlink[1]{}%
\providecommand \@@endlink[0]{}%
\providecommand \url  [0]{\begingroup\@sanitize@url \@url }%
\providecommand \@url [1]{\endgroup\@href {#1}{\urlprefix }}%
\providecommand \urlprefix  [0]{URL }%
\providecommand \Eprint [0]{\href }%
\providecommand \doibase [0]{https://doi.org/}%
\providecommand \selectlanguage [0]{\@gobble}%
\providecommand \bibinfo  [0]{\@secondoftwo}%
\providecommand \bibfield  [0]{\@secondoftwo}%
\providecommand \translation [1]{[#1]}%
\providecommand \BibitemOpen [0]{}%
\providecommand \bibitemStop [0]{}%
\providecommand \bibitemNoStop [0]{.\EOS\space}%
\providecommand \EOS [0]{\spacefactor3000\relax}%
\providecommand \BibitemShut  [1]{\csname bibitem#1\endcsname}%
\let\auto@bib@innerbib\@empty
\bibitem [{\citenamefont {Bogdanov}\ and\ \citenamefont
  {Yablonskii}(1989)}]{Bogdanov1989}%
  \BibitemOpen
  \bibfield  {author} {\bibinfo {author} {\bibfnamefont {A.}~\bibnamefont
  {Bogdanov}}\ and\ \bibinfo {author} {\bibfnamefont {D.}~\bibnamefont
  {Yablonskii}},\ }\bibfield  {title} {\bibinfo {title} {Thermodynamically
  stable "vortices" in magnetically ordered crystals. the mixed state of
  magnets},\ }\href@noop {} {\bibfield  {journal} {\bibinfo  {journal} {Sov.
  Phys. JETP}\ }\textbf {\bibinfo {volume} {68}},\ \bibinfo {pages} {101}
  (\bibinfo {year} {1989})}\BibitemShut {NoStop}%
\bibitem [{\citenamefont {Nagaosa}\ and\ \citenamefont
  {Tokura}(2013)}]{Nagaosa2013}%
  \BibitemOpen
  \bibfield  {author} {\bibinfo {author} {\bibfnamefont {N.}~\bibnamefont
  {Nagaosa}}\ and\ \bibinfo {author} {\bibfnamefont {Y.}~\bibnamefont
  {Tokura}},\ }\bibfield  {title} {\bibinfo {title} {Topological properties and
  dynamics of magnetic skyrmions},\ }\href
  {https://doi.org/10.1038/nnano.2013.243} {\bibfield  {journal} {\bibinfo
  {journal} {Nat. Nanotechnol.}\ }\textbf {\bibinfo {volume} {8}},\ \bibinfo
  {pages} {899} (\bibinfo {year} {2013})}\BibitemShut {NoStop}%
\bibitem [{\citenamefont {Kiselev}\ \emph {et~al.}(2011)\citenamefont
  {Kiselev}, \citenamefont {Bogdanov}, \citenamefont {Schäfer},\ and\
  \citenamefont {Rö{\ss}ler}}]{Kiselev2011}%
  \BibitemOpen
  \bibfield  {author} {\bibinfo {author} {\bibfnamefont {N.~S.}\ \bibnamefont
  {Kiselev}}, \bibinfo {author} {\bibfnamefont {A.~N.}\ \bibnamefont
  {Bogdanov}}, \bibinfo {author} {\bibfnamefont {R.}~\bibnamefont {Schäfer}},\
  and\ \bibinfo {author} {\bibfnamefont {U.~K.}\ \bibnamefont {Rö{\ss}ler}},\
  }\bibfield  {title} {\bibinfo {title} {Chiral skyrmions in thin magnetic
  films: new objects for magnetic storage technologies?},\ }\href
  {https://doi.org/10.1088/0022-3727/44/39/392001} {\bibfield  {journal}
  {\bibinfo  {journal} {J. Phys. D: Appl. Phys.}\ }\textbf {\bibinfo {volume}
  {44}},\ \bibinfo {pages} {392001} (\bibinfo {year} {2011})}\BibitemShut
  {NoStop}%
\bibitem [{\citenamefont {Fert}\ \emph {et~al.}(2017)\citenamefont {Fert},
  \citenamefont {Reyren},\ and\ \citenamefont {Cros}}]{Fert2017}%
  \BibitemOpen
  \bibfield  {author} {\bibinfo {author} {\bibfnamefont {A.}~\bibnamefont
  {Fert}}, \bibinfo {author} {\bibfnamefont {N.}~\bibnamefont {Reyren}},\ and\
  \bibinfo {author} {\bibfnamefont {V.}~\bibnamefont {Cros}},\ }\bibfield
  {title} {\bibinfo {title} {Magnetic skyrmions: advances in physics and
  potential applications},\ }\href {https://doi.org/10.1038/natrevmats.2017.31}
  {\bibfield  {journal} {\bibinfo  {journal} {Nat. Rev. Mater.}\ }\textbf
  {\bibinfo {volume} {2}},\ \bibinfo {pages} {17031} (\bibinfo {year}
  {2017})}\BibitemShut {NoStop}%
\bibitem [{\citenamefont {Back}\ \emph {et~al.}(2020)\citenamefont {Back},
  \citenamefont {Cros}, \citenamefont {Ebert}, \citenamefont {Everschor-Sitte},
  \citenamefont {Fert}, \citenamefont {Garst}, \citenamefont {Ma},
  \citenamefont {Mankovsky}, \citenamefont {Monchesky}, \citenamefont
  {Mostovoy}, \citenamefont {Nagaosa}, \citenamefont {Parkin}, \citenamefont
  {Pfleiderer}, \citenamefont {Reyren}, \citenamefont {Rosch}, \citenamefont
  {Taguchi}, \citenamefont {Tokura}, \citenamefont {von Bergmann},\ and\
  \citenamefont {Zang}}]{Back2020}%
  \BibitemOpen
  \bibfield  {author} {\bibinfo {author} {\bibfnamefont {C.}~\bibnamefont
  {Back}}, \bibinfo {author} {\bibfnamefont {V.}~\bibnamefont {Cros}}, \bibinfo
  {author} {\bibfnamefont {H.}~\bibnamefont {Ebert}}, \bibinfo {author}
  {\bibfnamefont {K.}~\bibnamefont {Everschor-Sitte}}, \bibinfo {author}
  {\bibfnamefont {A.}~\bibnamefont {Fert}}, \bibinfo {author} {\bibfnamefont
  {M.}~\bibnamefont {Garst}}, \bibinfo {author} {\bibfnamefont
  {T.}~\bibnamefont {Ma}}, \bibinfo {author} {\bibfnamefont {S.}~\bibnamefont
  {Mankovsky}}, \bibinfo {author} {\bibfnamefont {T.~L.}\ \bibnamefont
  {Monchesky}}, \bibinfo {author} {\bibfnamefont {M.}~\bibnamefont {Mostovoy}},
  \bibinfo {author} {\bibfnamefont {N.}~\bibnamefont {Nagaosa}}, \bibinfo
  {author} {\bibfnamefont {S.~S.~P.}\ \bibnamefont {Parkin}}, \bibinfo {author}
  {\bibfnamefont {C.}~\bibnamefont {Pfleiderer}}, \bibinfo {author}
  {\bibfnamefont {N.}~\bibnamefont {Reyren}}, \bibinfo {author} {\bibfnamefont
  {A.}~\bibnamefont {Rosch}}, \bibinfo {author} {\bibfnamefont
  {Y.}~\bibnamefont {Taguchi}}, \bibinfo {author} {\bibfnamefont
  {Y.}~\bibnamefont {Tokura}}, \bibinfo {author} {\bibfnamefont
  {K.}~\bibnamefont {von Bergmann}},\ and\ \bibinfo {author} {\bibfnamefont
  {J.}~\bibnamefont {Zang}},\ }\bibfield  {title} {\bibinfo {title} {The 2020
  skyrmionics roadmap},\ }\href {https://doi.org/10.1088/1361-6463/ab8418}
  {\bibfield  {journal} {\bibinfo  {journal} {J. Phys. D: Appl. Phys.}\
  }\textbf {\bibinfo {volume} {53}},\ \bibinfo {pages} {363001} (\bibinfo
  {year} {2020})}\BibitemShut {NoStop}%
\bibitem [{\citenamefont {Vedmedenko}\ and\ \citenamefont
  {Wiesendanger}(2019)}]{Vedmedenko2019}%
  \BibitemOpen
  \bibfield  {author} {\bibinfo {author} {\bibfnamefont {E.}~\bibnamefont
  {Vedmedenko}}\ and\ \bibinfo {author} {\bibfnamefont {R.}~\bibnamefont
  {Wiesendanger}},\ }\bibinfo {title} {Magnetic skyrmions on discrete
  lattices},\ in\ \href@noop {} {\emph {\bibinfo {booktitle} {Spintronic
  Handbook: Spin Transport and Magnetism}}},\ \bibinfo {editor} {edited by\
  \bibinfo {editor} {\bibfnamefont {I.~Z.}\ \bibnamefont {Evgeny Y.~Tsymbal}}}\
  (\bibinfo  {publisher} {CRC Press, Taylor \& Francis},\ \bibinfo {address}
  {Boca Raton},\ \bibinfo {year} {2019})\ Chap.~\bibinfo {chapter} {10}, pp.\
  \bibinfo {pages} {323--357},\ \bibinfo {edition} {2nd}\ ed.\BibitemShut
  {Stop}%
\bibitem [{\citenamefont {M\"uhlbauer}\ \emph {et~al.}(2009)\citenamefont
  {M\"uhlbauer}, \citenamefont {Binz}, \citenamefont {Jonietz}, \citenamefont
  {Pfleiderer}, \citenamefont {Rosch}, \citenamefont {Neubauer}, \citenamefont
  {Georgii},\ and\ \citenamefont {B\"oni}}]{Muhlbauer2009}%
  \BibitemOpen
  \bibfield  {author} {\bibinfo {author} {\bibfnamefont {S.}~\bibnamefont
  {M\"uhlbauer}}, \bibinfo {author} {\bibfnamefont {B.}~\bibnamefont {Binz}},
  \bibinfo {author} {\bibfnamefont {F.}~\bibnamefont {Jonietz}}, \bibinfo
  {author} {\bibfnamefont {C.}~\bibnamefont {Pfleiderer}}, \bibinfo {author}
  {\bibfnamefont {A.}~\bibnamefont {Rosch}}, \bibinfo {author} {\bibfnamefont
  {A.}~\bibnamefont {Neubauer}}, \bibinfo {author} {\bibfnamefont
  {R.}~\bibnamefont {Georgii}},\ and\ \bibinfo {author} {\bibfnamefont
  {P.}~\bibnamefont {B\"oni}},\ }\bibfield  {title} {\bibinfo {title} {Skyrmion
  lattice in a chiral magnet},\ }\href
  {https://doi.org/10.1126/science.1166767} {\bibfield  {journal} {\bibinfo
  {journal} {Science}\ }\textbf {\bibinfo {volume} {323}},\ \bibinfo {pages}
  {915} (\bibinfo {year} {2009})}\BibitemShut {NoStop}%
\bibitem [{\citenamefont {Yu}\ \emph {et~al.}(2010)\citenamefont {Yu},
  \citenamefont {Onose}, \citenamefont {Kanazawa}, \citenamefont {Park},
  \citenamefont {Han}, \citenamefont {Matsui}, \citenamefont {Nagaosa},\ and\
  \citenamefont {Tokura}}]{Yu2010}%
  \BibitemOpen
  \bibfield  {author} {\bibinfo {author} {\bibfnamefont {X.~Z.}\ \bibnamefont
  {Yu}}, \bibinfo {author} {\bibfnamefont {Y.}~\bibnamefont {Onose}}, \bibinfo
  {author} {\bibfnamefont {N.}~\bibnamefont {Kanazawa}}, \bibinfo {author}
  {\bibfnamefont {J.~H.}\ \bibnamefont {Park}}, \bibinfo {author}
  {\bibfnamefont {J.~H.}\ \bibnamefont {Han}}, \bibinfo {author} {\bibfnamefont
  {Y.}~\bibnamefont {Matsui}}, \bibinfo {author} {\bibfnamefont
  {N.}~\bibnamefont {Nagaosa}},\ and\ \bibinfo {author} {\bibfnamefont
  {Y.}~\bibnamefont {Tokura}},\ }\bibfield  {title} {\bibinfo {title}
  {Real-space observation of a two-dimensional skyrmion crystal},\ }\href
  {https://doi.org/10.1038/nature09124} {\bibfield  {journal} {\bibinfo
  {journal} {Nature}\ }\textbf {\bibinfo {volume} {465}},\ \bibinfo {pages}
  {901} (\bibinfo {year} {2010})}\BibitemShut {NoStop}%
\bibitem [{\citenamefont {Kotani}\ \emph {et~al.}(2018)\citenamefont {Kotani},
  \citenamefont {Harada}, \citenamefont {Malac}, \citenamefont {Salomons},
  \citenamefont {Hayashida},\ and\ \citenamefont {Mori}}]{Kotani2018}%
  \BibitemOpen
  \bibfield  {author} {\bibinfo {author} {\bibfnamefont {A.}~\bibnamefont
  {Kotani}}, \bibinfo {author} {\bibfnamefont {K.}~\bibnamefont {Harada}},
  \bibinfo {author} {\bibfnamefont {M.}~\bibnamefont {Malac}}, \bibinfo
  {author} {\bibfnamefont {M.}~\bibnamefont {Salomons}}, \bibinfo {author}
  {\bibfnamefont {M.}~\bibnamefont {Hayashida}},\ and\ \bibinfo {author}
  {\bibfnamefont {S.}~\bibnamefont {Mori}},\ }\bibfield  {title} {\bibinfo
  {title} {Observation of fege skyrmions by electron phase microscopy with
  hole-free phase plate},\ }\href {https://doi.org/10.1063/1.5028398}
  {\bibfield  {journal} {\bibinfo  {journal} {AIP Adv.}\ }\textbf {\bibinfo
  {volume} {8}},\ \bibinfo {pages} {055216} (\bibinfo {year}
  {2018})}\BibitemShut {NoStop}%
\bibitem [{\citenamefont {Romming}\ \emph {et~al.}(2013)\citenamefont
  {Romming}, \citenamefont {Hanneken}, \citenamefont {Menzel}, \citenamefont
  {Bickel}, \citenamefont {Wolter}, \citenamefont {von Bergmann}, \citenamefont
  {Kubetzka},\ and\ \citenamefont {Wiesendanger}}]{Romming2013}%
  \BibitemOpen
  \bibfield  {author} {\bibinfo {author} {\bibfnamefont {N.}~\bibnamefont
  {Romming}}, \bibinfo {author} {\bibfnamefont {C.}~\bibnamefont {Hanneken}},
  \bibinfo {author} {\bibfnamefont {M.}~\bibnamefont {Menzel}}, \bibinfo
  {author} {\bibfnamefont {J.~E.}\ \bibnamefont {Bickel}}, \bibinfo {author}
  {\bibfnamefont {B.}~\bibnamefont {Wolter}}, \bibinfo {author} {\bibfnamefont
  {K.}~\bibnamefont {von Bergmann}}, \bibinfo {author} {\bibfnamefont
  {A.}~\bibnamefont {Kubetzka}},\ and\ \bibinfo {author} {\bibfnamefont
  {R.}~\bibnamefont {Wiesendanger}},\ }\bibfield  {title} {\bibinfo {title}
  {Writing and deleting single magnetic skyrmions},\ }\href
  {https://doi.org/10.1126/science.1240573} {\bibfield  {journal} {\bibinfo
  {journal} {Science}\ }\textbf {\bibinfo {volume} {341}},\ \bibinfo {pages}
  {636} (\bibinfo {year} {2013})}\BibitemShut {NoStop}%
\bibitem [{\citenamefont {Hagemeister}\ \emph {et~al.}(2016)\citenamefont
  {Hagemeister}, \citenamefont {Iaia}, \citenamefont {Vedmedenko},
  \citenamefont {von Bergmann}, \citenamefont {Kubetzka},\ and\ \citenamefont
  {Wiesendanger}}]{Hagemeister2016}%
  \BibitemOpen
  \bibfield  {author} {\bibinfo {author} {\bibfnamefont {J.}~\bibnamefont
  {Hagemeister}}, \bibinfo {author} {\bibfnamefont {D.}~\bibnamefont {Iaia}},
  \bibinfo {author} {\bibfnamefont {E.~Y.}\ \bibnamefont {Vedmedenko}},
  \bibinfo {author} {\bibfnamefont {K.}~\bibnamefont {von Bergmann}}, \bibinfo
  {author} {\bibfnamefont {A.}~\bibnamefont {Kubetzka}},\ and\ \bibinfo
  {author} {\bibfnamefont {R.}~\bibnamefont {Wiesendanger}},\ }\bibfield
  {title} {\bibinfo {title} {Skyrmions at the edge: Confinement effects in
  fe/ir(111).},\ }\href
  {https://journals.aps.org/prl/abstract/10.1103/PhysRevLett.117.207202}
  {\bibfield  {journal} {\bibinfo  {journal} {Phys. Rev. Let}\ }\textbf
  {\bibinfo {volume} {117}},\ \bibinfo {pages} {207202} (\bibinfo {year}
  {2016})}\BibitemShut {NoStop}%
\bibitem [{\citenamefont {Vedmedenko}\ \emph {et~al.}(2020)\citenamefont
  {Vedmedenko}, \citenamefont {Kawakami}, \citenamefont {Sheka}, \citenamefont
  {Gambardella}, \citenamefont {Kirilyuk}, \citenamefont {Hirohata},
  \citenamefont {Binek}, \citenamefont {Chubykalo-Fesenko}, \citenamefont
  {Sanvito}, \citenamefont {Kirby}, \citenamefont {Grollier}, \citenamefont
  {Everschor-Sitte}, \citenamefont {Kampfrath}, \citenamefont {You},\ and\
  \citenamefont {Berger}}]{Vedmedenko2020}%
  \BibitemOpen
  \bibfield  {author} {\bibinfo {author} {\bibfnamefont {E.~Y.}\ \bibnamefont
  {Vedmedenko}}, \bibinfo {author} {\bibfnamefont {R.~K.}\ \bibnamefont
  {Kawakami}}, \bibinfo {author} {\bibfnamefont {D.~D.}\ \bibnamefont {Sheka}},
  \bibinfo {author} {\bibfnamefont {P.}~\bibnamefont {Gambardella}}, \bibinfo
  {author} {\bibfnamefont {A.}~\bibnamefont {Kirilyuk}}, \bibinfo {author}
  {\bibfnamefont {A.}~\bibnamefont {Hirohata}}, \bibinfo {author}
  {\bibfnamefont {C.}~\bibnamefont {Binek}}, \bibinfo {author} {\bibfnamefont
  {O.}~\bibnamefont {Chubykalo-Fesenko}}, \bibinfo {author} {\bibfnamefont
  {S.}~\bibnamefont {Sanvito}}, \bibinfo {author} {\bibfnamefont {B.~J.}\
  \bibnamefont {Kirby}}, \bibinfo {author} {\bibfnamefont {J.}~\bibnamefont
  {Grollier}}, \bibinfo {author} {\bibfnamefont {K.}~\bibnamefont
  {Everschor-Sitte}}, \bibinfo {author} {\bibfnamefont {T.}~\bibnamefont
  {Kampfrath}}, \bibinfo {author} {\bibfnamefont {C.-Y.}\ \bibnamefont {You}},\
  and\ \bibinfo {author} {\bibfnamefont {A.}~\bibnamefont {Berger}},\
  }\bibfield  {title} {\bibinfo {title} {The 2020 magnetism roadmap},\ }\href
  {https://doi.org/10.1088/1361-6463/ab9d98} {\bibfield  {journal} {\bibinfo
  {journal} {J. Phys. D: App. Phys.}\ }\textbf {\bibinfo {volume} {53}},\
  \bibinfo {pages} {453001} (\bibinfo {year} {2020})}\BibitemShut {NoStop}%
\bibitem [{\citenamefont {Heinze}\ \emph {et~al.}(2011)\citenamefont {Heinze},
  \citenamefont {von Bergmann}, \citenamefont {Menzel}, \citenamefont {Brede},
  \citenamefont {Kubetzka}, \citenamefont {Wiesendanger}, \citenamefont
  {Bihlmayer},\ and\ \citenamefont {Blügel}}]{Heinze2011}%
  \BibitemOpen
  \bibfield  {author} {\bibinfo {author} {\bibfnamefont {S.}~\bibnamefont
  {Heinze}}, \bibinfo {author} {\bibfnamefont {K.}~\bibnamefont {von
  Bergmann}}, \bibinfo {author} {\bibfnamefont {M.}~\bibnamefont {Menzel}},
  \bibinfo {author} {\bibfnamefont {J.}~\bibnamefont {Brede}}, \bibinfo
  {author} {\bibfnamefont {A.}~\bibnamefont {Kubetzka}}, \bibinfo {author}
  {\bibfnamefont {R.}~\bibnamefont {Wiesendanger}}, \bibinfo {author}
  {\bibfnamefont {G.}~\bibnamefont {Bihlmayer}},\ and\ \bibinfo {author}
  {\bibfnamefont {S.}~\bibnamefont {Blügel}},\ }\bibfield  {title} {\bibinfo
  {title} {Spontaneous atomic-scale magnetic skyrmion lattice in two
  dimensions},\ }\href {https://doi.org/10.1038/nphys2045} {\bibfield
  {journal} {\bibinfo  {journal} {Nat. Phys.}\ }\textbf {\bibinfo {volume}
  {7}},\ \bibinfo {pages} {713} (\bibinfo {year} {2011})}\BibitemShut {NoStop}%
\bibitem [{\citenamefont {Sampaio}\ \emph {et~al.}(2013)\citenamefont
  {Sampaio}, \citenamefont {Cros}, \citenamefont {Rohart}, \citenamefont
  {Thiaville},\ and\ \citenamefont {Fert}}]{Sampaio2013}%
  \BibitemOpen
  \bibfield  {author} {\bibinfo {author} {\bibfnamefont {J.}~\bibnamefont
  {Sampaio}}, \bibinfo {author} {\bibfnamefont {V.}~\bibnamefont {Cros}},
  \bibinfo {author} {\bibfnamefont {S.}~\bibnamefont {Rohart}}, \bibinfo
  {author} {\bibfnamefont {A.}~\bibnamefont {Thiaville}},\ and\ \bibinfo
  {author} {\bibfnamefont {A.}~\bibnamefont {Fert}},\ }\bibfield  {title}
  {\bibinfo {title} {Nucleation, stability and current-induced motion of
  isolated magnetic skyrmions in nanostructures},\ }\href
  {https://doi.org/10.1038/nnano.2013.210} {\bibfield  {journal} {\bibinfo
  {journal} {Nat. Nanotechnol.}\ }\textbf {\bibinfo {volume} {8}},\ \bibinfo
  {pages} {839} (\bibinfo {year} {2013})}\BibitemShut {NoStop}%
\bibitem [{\citenamefont {Yuan}\ and\ \citenamefont {Wang}(2016)}]{Yuan2016}%
  \BibitemOpen
  \bibfield  {author} {\bibinfo {author} {\bibfnamefont {H.~Y.}\ \bibnamefont
  {Yuan}}\ and\ \bibinfo {author} {\bibfnamefont {X.~R.}\ \bibnamefont
  {Wang}},\ }\bibfield  {title} {\bibinfo {title} {Skyrmion creation and
  manipulation by nano-second current pulses.},\ }\href@noop {} {\bibfield
  {journal} {\bibinfo  {journal} {Sci. Rep.}\ }\textbf {\bibinfo {volume}
  {6}},\ \bibinfo {pages} {22638} (\bibinfo {year} {2016})}\BibitemShut
  {NoStop}%
\bibitem [{\citenamefont {Stier}\ \emph {et~al.}(2017)\citenamefont {Stier},
  \citenamefont {H\"ausler}, \citenamefont {Posske}, \citenamefont {Gurski},\
  and\ \citenamefont {Thorwart}}]{Stier2017}%
  \BibitemOpen
  \bibfield  {author} {\bibinfo {author} {\bibfnamefont {M.}~\bibnamefont
  {Stier}}, \bibinfo {author} {\bibfnamefont {W.}~\bibnamefont {H\"ausler}},
  \bibinfo {author} {\bibfnamefont {T.}~\bibnamefont {Posske}}, \bibinfo
  {author} {\bibfnamefont {G.}~\bibnamefont {Gurski}},\ and\ \bibinfo {author}
  {\bibfnamefont {M.}~\bibnamefont {Thorwart}},\ }\bibfield  {title} {\bibinfo
  {title} {Skyrmion--anti-skyrmion pair creation by in-plane currents},\ }\href
  {https://doi.org/10.1103/PhysRevLett.118.267203} {\bibfield  {journal}
  {\bibinfo  {journal} {Phys. Rev. Lett.}\ }\textbf {\bibinfo {volume} {118}},\
  \bibinfo {pages} {267203} (\bibinfo {year} {2017})}\BibitemShut {NoStop}%
\bibitem [{\citenamefont {Flovik}\ \emph {et~al.}(2017)\citenamefont {Flovik},
  \citenamefont {Qaiumzadeh}, \citenamefont {Nandy}, \citenamefont {Heo},\ and\
  \citenamefont {Rasing}}]{Flovik2017}%
  \BibitemOpen
  \bibfield  {author} {\bibinfo {author} {\bibfnamefont {V.}~\bibnamefont
  {Flovik}}, \bibinfo {author} {\bibfnamefont {A.}~\bibnamefont {Qaiumzadeh}},
  \bibinfo {author} {\bibfnamefont {A.~K.}\ \bibnamefont {Nandy}}, \bibinfo
  {author} {\bibfnamefont {C.}~\bibnamefont {Heo}},\ and\ \bibinfo {author}
  {\bibfnamefont {T.}~\bibnamefont {Rasing}},\ }\bibfield  {title} {\bibinfo
  {title} {Generation of single skyrmions by picosecond magnetic field
  pulses},\ }\href {https://doi.org/10.1103/PhysRevB.96.140411} {\bibfield
  {journal} {\bibinfo  {journal} {Phys. Rev. B}\ }\textbf {\bibinfo {volume}
  {96}},\ \bibinfo {pages} {140411} (\bibinfo {year} {2017})}\BibitemShut
  {NoStop}%
\bibitem [{\citenamefont {Schäffer}\ \emph {et~al.}(2020)\citenamefont
  {Schäffer}, \citenamefont {Siegl}, \citenamefont {Stier}, \citenamefont
  {Posske}, \citenamefont {Berakdar}, \citenamefont {Thorwart}, \citenamefont
  {Wiesendanger},\ and\ \citenamefont {Vedmedenko}}]{Schaffer2020}%
  \BibitemOpen
  \bibfield  {author} {\bibinfo {author} {\bibfnamefont {A.~F.}\ \bibnamefont
  {Schäffer}}, \bibinfo {author} {\bibfnamefont {P.}~\bibnamefont {Siegl}},
  \bibinfo {author} {\bibfnamefont {M.}~\bibnamefont {Stier}}, \bibinfo
  {author} {\bibfnamefont {T.}~\bibnamefont {Posske}}, \bibinfo {author}
  {\bibfnamefont {J.}~\bibnamefont {Berakdar}}, \bibinfo {author}
  {\bibfnamefont {M.}~\bibnamefont {Thorwart}}, \bibinfo {author}
  {\bibfnamefont {R.}~\bibnamefont {Wiesendanger}},\ and\ \bibinfo {author}
  {\bibfnamefont {E.~Y.}\ \bibnamefont {Vedmedenko}},\ }\bibfield  {title}
  {\bibinfo {title} {Rotating edge-field driven processing of chiral spin
  textures in racetrack devices},\ }\href
  {https://doi.org/10.1038/s41598-020-77337-y} {\bibfield  {journal} {\bibinfo
  {journal} {Sci. Rep.}\ }\textbf {\bibinfo {volume} {10}},\ \bibinfo {pages}
  {20400} (\bibinfo {year} {2020})}\BibitemShut {NoStop}%
\bibitem [{\citenamefont {Raeliarijaona}\ \emph {et~al.}(2018)\citenamefont
  {Raeliarijaona}, \citenamefont {Nepal},\ and\ \citenamefont
  {Kovalev}}]{Raeliarijaona2018}%
  \BibitemOpen
  \bibfield  {author} {\bibinfo {author} {\bibfnamefont {A.}~\bibnamefont
  {Raeliarijaona}}, \bibinfo {author} {\bibfnamefont {R.}~\bibnamefont
  {Nepal}},\ and\ \bibinfo {author} {\bibfnamefont {A.~A.}\ \bibnamefont
  {Kovalev}},\ }\bibfield  {title} {\bibinfo {title} {Boundary twists,
  instabilities, and creation of skyrmions and antiskyrmions},\ }\href
  {https://doi.org/10.1103/PhysRevMaterials.2.124401} {\bibfield  {journal}
  {\bibinfo  {journal} {Phys. Rev. Mater.}\ }\textbf {\bibinfo {volume} {2}},\
  \bibinfo {pages} {124401} (\bibinfo {year} {2018})}\BibitemShut {NoStop}%
\bibitem [{\citenamefont {Hagemeister}\ \emph {et~al.}(2015)\citenamefont
  {Hagemeister}, \citenamefont {Romming}, \citenamefont {von Bergmann},
  \citenamefont {Vedmedenko},\ and\ \citenamefont
  {Wiesendanger}}]{Hagemeister2015}%
  \BibitemOpen
  \bibfield  {author} {\bibinfo {author} {\bibfnamefont {J.}~\bibnamefont
  {Hagemeister}}, \bibinfo {author} {\bibfnamefont {N.}~\bibnamefont
  {Romming}}, \bibinfo {author} {\bibfnamefont {K.}~\bibnamefont {von
  Bergmann}}, \bibinfo {author} {\bibfnamefont {E.~Y.}\ \bibnamefont
  {Vedmedenko}},\ and\ \bibinfo {author} {\bibfnamefont {R.}~\bibnamefont
  {Wiesendanger}},\ }\bibfield  {title} {\bibinfo {title} {Stability of single
  skyrmionic bits},\ }\href {https://doi.org/10.1038/ncomms9455} {\bibfield
  {journal} {\bibinfo  {journal} {Nat. Commun.}\ }\textbf {\bibinfo {volume}
  {6}},\ \bibinfo {pages} {8455} (\bibinfo {year} {2015})}\BibitemShut
  {NoStop}%
\bibitem [{\citenamefont {Siemens}\ \emph {et~al.}(2016)\citenamefont
  {Siemens}, \citenamefont {Zhang}, \citenamefont {Hagemeister}, \citenamefont
  {Vedmedenko},\ and\ \citenamefont {Wiesendanger}}]{Siemens2016}%
  \BibitemOpen
  \bibfield  {author} {\bibinfo {author} {\bibfnamefont {A.}~\bibnamefont
  {Siemens}}, \bibinfo {author} {\bibfnamefont {Y.}~\bibnamefont {Zhang}},
  \bibinfo {author} {\bibfnamefont {J.}~\bibnamefont {Hagemeister}}, \bibinfo
  {author} {\bibfnamefont {E.~Y.}\ \bibnamefont {Vedmedenko}},\ and\ \bibinfo
  {author} {\bibfnamefont {R.}~\bibnamefont {Wiesendanger}},\ }\bibfield
  {title} {\bibinfo {title} {Minimal radius of magnetic skyrmions: Statics and
  dynamics},\ }\href
  {https://iopscience.iop.org/article/10.1088/1367-2630/18/4/045021} {\bibfield
   {journal} {\bibinfo  {journal} {New J. Phys.}\ } (\bibinfo {year}
  {2016})}\BibitemShut {NoStop}%
\bibitem [{\citenamefont {Cortés-Ortuño}\ \emph {et~al.}(2017)\citenamefont
  {Cortés-Ortuño}, \citenamefont {Wang}, \citenamefont {Beg}, \citenamefont
  {Pepper}, \citenamefont {Bisotti}, \citenamefont {Carey}, \citenamefont
  {Vousden}, \citenamefont {Kluyver}, \citenamefont {Hovorka},\ and\
  \citenamefont {Fangohr}}]{CortesOrtuno2017}%
  \BibitemOpen
  \bibfield  {author} {\bibinfo {author} {\bibfnamefont {D.}~\bibnamefont
  {Cortés-Ortuño}}, \bibinfo {author} {\bibfnamefont {W.}~\bibnamefont
  {Wang}}, \bibinfo {author} {\bibfnamefont {M.}~\bibnamefont {Beg}}, \bibinfo
  {author} {\bibfnamefont {R.~A.}\ \bibnamefont {Pepper}}, \bibinfo {author}
  {\bibfnamefont {M.-A.}\ \bibnamefont {Bisotti}}, \bibinfo {author}
  {\bibfnamefont {R.}~\bibnamefont {Carey}}, \bibinfo {author} {\bibfnamefont
  {M.}~\bibnamefont {Vousden}}, \bibinfo {author} {\bibfnamefont
  {T.}~\bibnamefont {Kluyver}}, \bibinfo {author} {\bibfnamefont
  {O.}~\bibnamefont {Hovorka}},\ and\ \bibinfo {author} {\bibfnamefont
  {H.}~\bibnamefont {Fangohr}},\ }\bibfield  {title} {\bibinfo {title} {Thermal
  stability and topological protection of skyrmions in nanotracks},\ }\href
  {https://doi.org/10.1038/s41598-017-03391-8} {\bibfield  {journal} {\bibinfo
  {journal} {Sci. Rep.}\ }\textbf {\bibinfo {volume} {7}},\ \bibinfo {pages}
  {4060} (\bibinfo {year} {2017})}\BibitemShut {NoStop}%
\bibitem [{\citenamefont {Rold\'an-Molina}\ \emph {et~al.}(2015)\citenamefont
  {Rold\'an-Molina}, \citenamefont {Santander}, \citenamefont {Nunez},\ and\
  \citenamefont {Fern\'andez-Rossier}}]{RoldanMolina2015}%
  \BibitemOpen
  \bibfield  {author} {\bibinfo {author} {\bibfnamefont {A.}~\bibnamefont
  {Rold\'an-Molina}}, \bibinfo {author} {\bibfnamefont {M.~J.}\ \bibnamefont
  {Santander}}, \bibinfo {author} {\bibfnamefont {A.~S.}\ \bibnamefont
  {Nunez}},\ and\ \bibinfo {author} {\bibfnamefont {J.}~\bibnamefont
  {Fern\'andez-Rossier}},\ }\bibfield  {title} {\bibinfo {title} {Quantum
  fluctuations stabilize skyrmion textures},\ }\href
  {https://doi.org/10.1103/PhysRevB.92.245436} {\bibfield  {journal} {\bibinfo
  {journal} {Phys. Rev. B}\ }\textbf {\bibinfo {volume} {92}},\ \bibinfo
  {pages} {245436} (\bibinfo {year} {2015})}\BibitemShut {NoStop}%
\bibitem [{\citenamefont {Derras-Chouk}\ \emph {et~al.}(2018)\citenamefont
  {Derras-Chouk}, \citenamefont {Chudnovsky},\ and\ \citenamefont
  {Garanin}}]{DerrasChouk2018}%
  \BibitemOpen
  \bibfield  {author} {\bibinfo {author} {\bibfnamefont {A.}~\bibnamefont
  {Derras-Chouk}}, \bibinfo {author} {\bibfnamefont {E.~M.}\ \bibnamefont
  {Chudnovsky}},\ and\ \bibinfo {author} {\bibfnamefont {D.~A.}\ \bibnamefont
  {Garanin}},\ }\bibfield  {title} {\bibinfo {title} {Quantum collapse of a
  magnetic skyrmion},\ }\href {https://doi.org/10.1103/PhysRevB.98.024423}
  {\bibfield  {journal} {\bibinfo  {journal} {Phys. Rev. B}\ }\textbf {\bibinfo
  {volume} {98}},\ \bibinfo {pages} {024423} (\bibinfo {year}
  {2018})}\BibitemShut {NoStop}%
\bibitem [{\citenamefont {Sotnikov}\ \emph {et~al.}(2021)\citenamefont
  {Sotnikov}, \citenamefont {Mazurenko}, \citenamefont {Colbois}, \citenamefont
  {Mila}, \citenamefont {Katsnelson},\ and\ \citenamefont
  {Stepanov}}]{Sotnikov2021}%
  \BibitemOpen
  \bibfield  {author} {\bibinfo {author} {\bibfnamefont {O.~M.}\ \bibnamefont
  {Sotnikov}}, \bibinfo {author} {\bibfnamefont {V.~V.}\ \bibnamefont
  {Mazurenko}}, \bibinfo {author} {\bibfnamefont {J.}~\bibnamefont {Colbois}},
  \bibinfo {author} {\bibfnamefont {F.}~\bibnamefont {Mila}}, \bibinfo {author}
  {\bibfnamefont {M.~I.}\ \bibnamefont {Katsnelson}},\ and\ \bibinfo {author}
  {\bibfnamefont {E.~A.}\ \bibnamefont {Stepanov}},\ }\bibfield  {title}
  {\bibinfo {title} {Probing the topology of the quantum analog of a classical
  skyrmion},\ }\href {https://doi.org/10.1103/PhysRevB.103.L060404} {\bibfield
  {journal} {\bibinfo  {journal} {Phys. Rev. B}\ }\textbf {\bibinfo {volume}
  {103}},\ \bibinfo {pages} {L060404} (\bibinfo {year} {2021})}\BibitemShut
  {NoStop}%
\bibitem [{\citenamefont {Gauyacq}\ and\ \citenamefont
  {Lorente}(2019)}]{Gauyacq2019a}%
  \BibitemOpen
  \bibfield  {author} {\bibinfo {author} {\bibfnamefont {J.-P.}\ \bibnamefont
  {Gauyacq}}\ and\ \bibinfo {author} {\bibfnamefont {N.}~\bibnamefont
  {Lorente}},\ }\bibfield  {title} {\bibinfo {title} {A model for individual
  quantal nano-skyrmions},\ }\href
  {https://iopscience.iop.org/article/10.1088/1361-648X/ab1f3a/pdf} {\bibfield
  {journal} {\bibinfo  {journal} {J. Phys.: Condens. Matter}\ }\textbf
  {\bibinfo {volume} {31}},\ \bibinfo {pages} {335001} (\bibinfo {year}
  {2019})}\BibitemShut {NoStop}%
\bibitem [{\citenamefont {Lohani}\ \emph {et~al.}(2019)\citenamefont {Lohani},
  \citenamefont {Hickey}, \citenamefont {Masell},\ and\ \citenamefont
  {Rosch}}]{Lohani2019a}%
  \BibitemOpen
  \bibfield  {author} {\bibinfo {author} {\bibfnamefont {V.}~\bibnamefont
  {Lohani}}, \bibinfo {author} {\bibfnamefont {C.}~\bibnamefont {Hickey}},
  \bibinfo {author} {\bibfnamefont {J.}~\bibnamefont {Masell}},\ and\ \bibinfo
  {author} {\bibfnamefont {A.}~\bibnamefont {Rosch}},\ }\bibfield  {title}
  {\bibinfo {title} {Quantum skyrmions in frustrated ferromagnets},\ }\href
  {https://doi.org/10.1103/PhysRevX.9.041063} {\bibfield  {journal} {\bibinfo
  {journal} {Phys. Rev. X}\ }\textbf {\bibinfo {volume} {9}},\ \bibinfo {eid}
  {041063} (\bibinfo {year} {2019})}\BibitemShut {NoStop}%
\bibitem [{\citenamefont {Berg}\ and\ \citenamefont
  {Lüscher}(1981)}]{Berg1981}%
  \BibitemOpen
  \bibfield  {author} {\bibinfo {author} {\bibfnamefont {B.}~\bibnamefont
  {Berg}}\ and\ \bibinfo {author} {\bibfnamefont {M.}~\bibnamefont
  {Lüscher}},\ }\bibfield  {title} {\bibinfo {title} {Definition and
  statistical distributions of a topological number in the lattice o(3)
  $\sigma$-model},\ }\href
  {https://doi.org/https://doi.org/10.1016/0550-3213(81)90568-X} {\bibfield
  {journal} {\bibinfo  {journal} {Nucl. Phys. B}\ }\textbf {\bibinfo {volume}
  {190}},\ \bibinfo {pages} {412} (\bibinfo {year} {1981})}\BibitemShut
  {NoStop}%
\bibitem [{\citenamefont {Vedmedenko}\ and\ \citenamefont
  {Altwein}(2014)}]{Vedmedenko2014}%
  \BibitemOpen
  \bibfield  {author} {\bibinfo {author} {\bibfnamefont {E.~Y.}\ \bibnamefont
  {Vedmedenko}}\ and\ \bibinfo {author} {\bibfnamefont {D.}~\bibnamefont
  {Altwein}},\ }\bibfield  {title} {\bibinfo {title} {Topologically protected
  magnetic helix for all-spin-based applications},\ }\href
  {https://doi.org/10.1103/PhysRevLett.112.017206} {\bibfield  {journal}
  {\bibinfo  {journal} {Phys. Rev. Lett.}\ }\textbf {\bibinfo {volume} {112}},\
  \bibinfo {pages} {017206} (\bibinfo {year} {2014})}\BibitemShut {NoStop}%
\bibitem [{\citenamefont {Posske}\ and\ \citenamefont
  {Thorwart}(2019)}]{Posske2019}%
  \BibitemOpen
  \bibfield  {author} {\bibinfo {author} {\bibfnamefont {T.}~\bibnamefont
  {Posske}}\ and\ \bibinfo {author} {\bibfnamefont {M.}~\bibnamefont
  {Thorwart}},\ }\bibfield  {title} {\bibinfo {title} {Winding up quantum spin
  helices: How avoided level crossings exile classical topological
  protection},\ }\href {https://doi.org/10.1103/PhysRevLett.122.097204}
  {\bibfield  {journal} {\bibinfo  {journal} {Phys. Rev. Lett.}\ }\textbf
  {\bibinfo {volume} {122}},\ \bibinfo {pages} {097204} (\bibinfo {year}
  {2019})}\BibitemShut {NoStop}%
\bibitem [{\citenamefont {Kato}(1950)}]{Kato1950}%
  \BibitemOpen
  \bibfield  {author} {\bibinfo {author} {\bibfnamefont {T.}~\bibnamefont
  {Kato}},\ }\bibfield  {title} {\bibinfo {title} {On the adiabatic theorem of
  quantum mechanics},\ }\href {https://doi.org/10.1143/JPSJ.5.435} {\bibfield
  {journal} {\bibinfo  {journal} {J. Phys. Soc. Japan}\ }\textbf {\bibinfo
  {volume} {5}},\ \bibinfo {pages} {435} (\bibinfo {year} {1950})}\BibitemShut
  {NoStop}%
\bibitem [{\citenamefont {Spethmann}\ \emph {et~al.}(2022)\citenamefont
  {Spethmann}, \citenamefont {Vedmedenko}, \citenamefont {Wiesendanger},
  \citenamefont {Kubetzka},\ and\ \citenamefont {von
  Bergmann}}]{Spethmann2022}%
  \BibitemOpen
  \bibfield  {author} {\bibinfo {author} {\bibfnamefont {J.}~\bibnamefont
  {Spethmann}}, \bibinfo {author} {\bibfnamefont {E.~Y.}\ \bibnamefont
  {Vedmedenko}}, \bibinfo {author} {\bibfnamefont {R.}~\bibnamefont
  {Wiesendanger}}, \bibinfo {author} {\bibfnamefont {A.}~\bibnamefont
  {Kubetzka}},\ and\ \bibinfo {author} {\bibfnamefont {K.}~\bibnamefont {von
  Bergmann}},\ }\bibfield  {title} {\bibinfo {title} {Zero-field skyrmionic
  states and in-field edge-skyrmions induced by boundary tuning},\ }\href
  {https://doi.org/10.1038/s42005-021-00796-w} {\bibfield  {journal} {\bibinfo
  {journal} {Communications Physics}\ }\textbf {\bibinfo {volume} {5}},\
  \bibinfo {pages} {19} (\bibinfo {year} {2022})}\BibitemShut {NoStop}%
\bibitem [{\citenamefont {Ryu}\ \emph {et~al.}(2010)\citenamefont {Ryu},
  \citenamefont {Schnyder}, \citenamefont {Furusaki},\ and\ \citenamefont
  {Ludwig}}]{Ryu2010}%
  \BibitemOpen
  \bibfield  {author} {\bibinfo {author} {\bibfnamefont {S.}~\bibnamefont
  {Ryu}}, \bibinfo {author} {\bibfnamefont {A.~P.}\ \bibnamefont {Schnyder}},
  \bibinfo {author} {\bibfnamefont {A.}~\bibnamefont {Furusaki}},\ and\
  \bibinfo {author} {\bibfnamefont {A.~W.~W.}\ \bibnamefont {Ludwig}},\
  }\bibfield  {title} {\bibinfo {title} {Topological insulators and
  superconductors: tenfold way and dimensional hierarchy},\ }\href
  {https://doi.org/10.1088/1367-2630/12/6/065010} {\bibfield  {journal}
  {\bibinfo  {journal} {New J. Phys.}\ }\textbf {\bibinfo {volume} {12}},\
  \bibinfo {pages} {065010} (\bibinfo {year} {2010})}\BibitemShut {NoStop}%
\bibitem [{\citenamefont {Metropolis}\ \emph {et~al.}(1953)\citenamefont
  {Metropolis}, \citenamefont {Rosenbluth}, \citenamefont {Rosenbluth},
  \citenamefont {Teller},\ and\ \citenamefont {Teller}}]{Metropolis1953}%
  \BibitemOpen
  \bibfield  {author} {\bibinfo {author} {\bibfnamefont {N.}~\bibnamefont
  {Metropolis}}, \bibinfo {author} {\bibfnamefont {A.~W.}\ \bibnamefont
  {Rosenbluth}}, \bibinfo {author} {\bibfnamefont {M.~N.}\ \bibnamefont
  {Rosenbluth}}, \bibinfo {author} {\bibfnamefont {A.~H.}\ \bibnamefont
  {Teller}},\ and\ \bibinfo {author} {\bibfnamefont {E.}~\bibnamefont
  {Teller}},\ }\bibfield  {title} {\bibinfo {title} {Equation of state
  calculations by fast computing machines},\ }\href
  {https://doi.org/10.1063/1.1699114} {\bibfield  {journal} {\bibinfo
  {journal} {J. Chem. Phys.}\ }\textbf {\bibinfo {volume} {21}},\ \bibinfo
  {pages} {1087} (\bibinfo {year} {1953})}\BibitemShut {NoStop}%
\end{thebibliography}
%

\appendix
\clearpage
\onecolumngrid
\beginsupplement
\section{Appendix: Methods}
\subsection{Adiabatic evolution}
An adiabatic process is a change of a Hamiltonian that is infinitely slow such that, if the quantum system is initially in an instantaneous eigenstate of the system, it remains in the corresponding instantaneous eigenstate over the course of changing the Hamiltonian.
In the main text, we study the adiabatic evolution of the system's ground state. 
The method we apply is based on the adiabatic evolution operator $K(s)$ introduced by Kato in Ref. \cite{Kato1950}.
$K(s)$ adiabatically propagates an initial state $\ket{\phi_n(0)}$ to $\ket{\phi_n(s)}$, where $s$ is a parameterized time ranging from $0$ to $1$.
For infinitesimal time steps $\delta$, $K(s)$ can be expressed in terms of projectors $P(s)=\ket{\phi_n(s)}\bra{\phi_n(s)}$ of the instantaneous eigenstates as
\begin{equation}\label{AdaiabticEvol}
K(s)=\lim_{\delta \to 0}\prod_{n=0}^{s/\delta}P(n\cdot\delta).
\end{equation}
This allows us to simulate the adiabatic evolution by consecutively applying projectors to $\ket{\phi_n(0)}$.
In our simulations, we consider a discrete step size $\epsilon$ instead of infinitely small steps. 
We consider the overlap of two quantum states $\ket{\phi_n}$ and $\ket{\phi_m}$ as $S_{n,m} = \braket{\phi_n|\phi_m}$, to identify on which instantaneous eigenstate, $\ket{\phi_n(0)}$ needs to be projected.
In the limit $\epsilon\rightarrow 0$, the overlap of the instantaneous state and its adiabatic successor is always $1$ while it is $0$ with all the other states.
For finite $\epsilon$, the overlap in general takes values between $0$ and $1$. For our simulations, we chose $\epsilon$ such that the overlap of the ground state is $>0.98$. That corresponds to an $\epsilon$ between $0.1$ and $0.01$, depending on the interaction parameters.  We chose the eigenstate with the largest overlap as the adiabatic successor.
If a degeneracy occurs, which we numerically assume when the energy difference between two states is below $10^{-3}J$, the projection is performed on the degenerate subspace.
\subsection{\texorpdfstring{Calculation of the topological phase boundary for classical $3\times 3$ and $4\times 4$ spin lattices}{Calculation of the topological phase boundary}}
In order to understand the peculiarity of the quantum skyrmionic ground states with $D \approx 0$ shown in Fig.~1 in the main text, we resolve the parameter boundary between skyrmionic and ferromagnetic ground states in $3 \times 3$ and $4 \times 4$ lattices of classical spins as well. To do so, we implement the Hamiltonian in Eq.~(1) of the main text for classical spins of a fixed magnitude, determine the ground state by a zero-temperature Metropolis algorithm \cite{Metropolis1953}, and calculate the corresponding the topological index $C$ in dependence on the DMI $D$ and the Heisenberg anisotropy $\Delta$. Within the zero-temperature Metropolis algorithm, we start with three different initial states: a random spin configuration, a ferromagnetic spin configuration pointing in the $(0,0,-1)$-direction, and a skyrmionic configuration as shown in Fig.~1 (b) in the main text. In each iteration, we change all spins uniformly randomly on the unit sphere with maximal absolute change of 0.02 in each direction. We can use the unit sphere here instead of a sphere with radius $\hbar/2$ because a global factor only rescales the energies, leaving the ground state characteristics unchanged. We evolve the algorithm for $s=10000$ steps and ensure its convergence by having a negligible change of the average energy well before $s$ iterations are reached. The local minima that are reached are compared and the one with minimal energy is regarded as the systems ground state. To resolve the change in the topological number, we perform a binary search in $D$ for $\Delta = 0,0.1,0.2,...,1.2$ with seven steps, resulting in an error bar of 0.015625 in $D$-direction.
\section{Appendix: Different Rotation Schemes}
\textit{Four-Edge Rotation.--}
\begin{figure*}
	\begin{minipage}{0.32\textwidth}
		\begin{picture}(0,0)
		\put(-55,2){{\large (a)}}
		\end{picture}
		\includegraphics[width=\linewidth,height=\textheight,keepaspectratio]{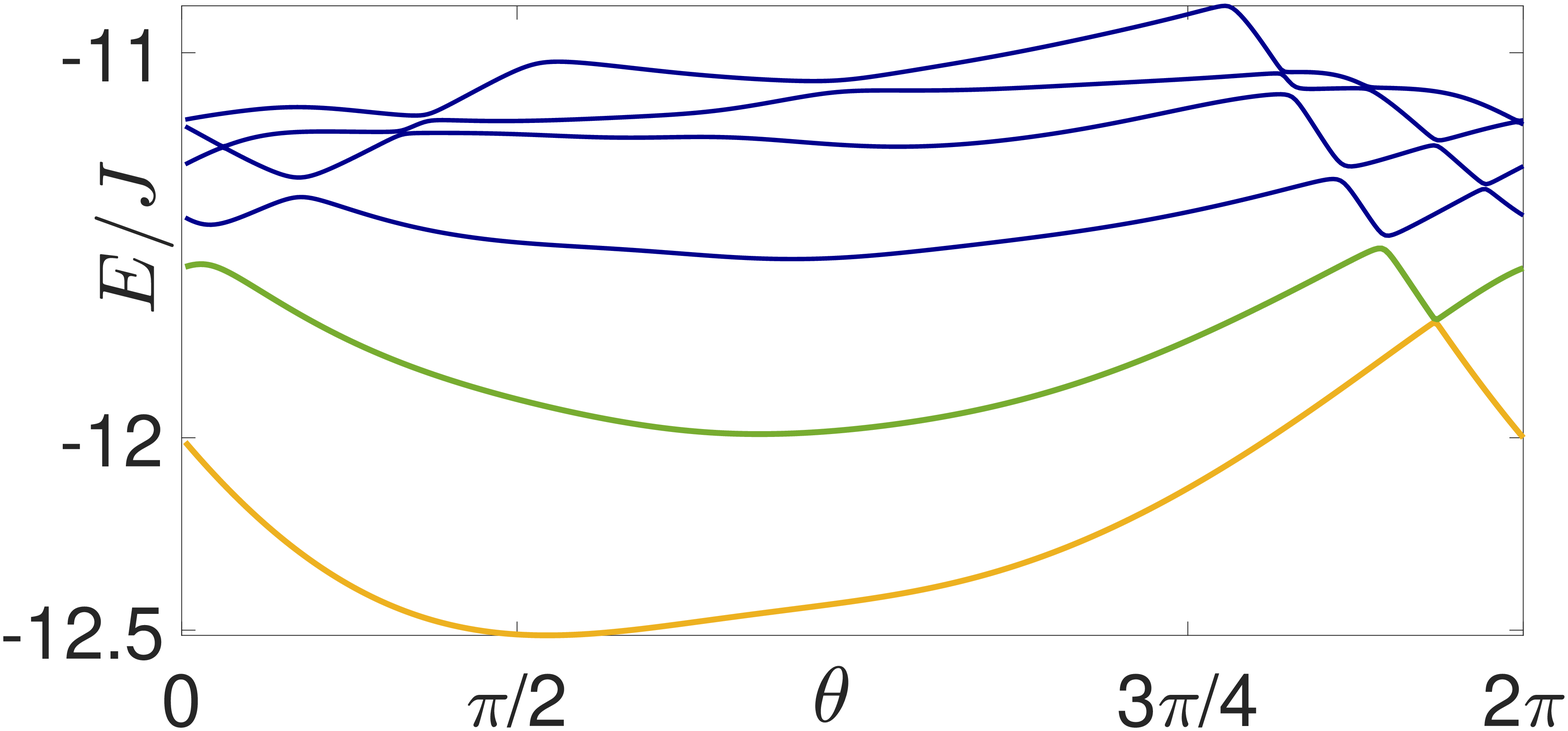}
	\end{minipage}
	\begin{minipage}{0.32\textwidth}
		\begin{picture}(0,0)
		\put(-55,2){{\large (b)}}
		\end{picture}
		\includegraphics[width=\linewidth, height=\textheight,keepaspectratio]{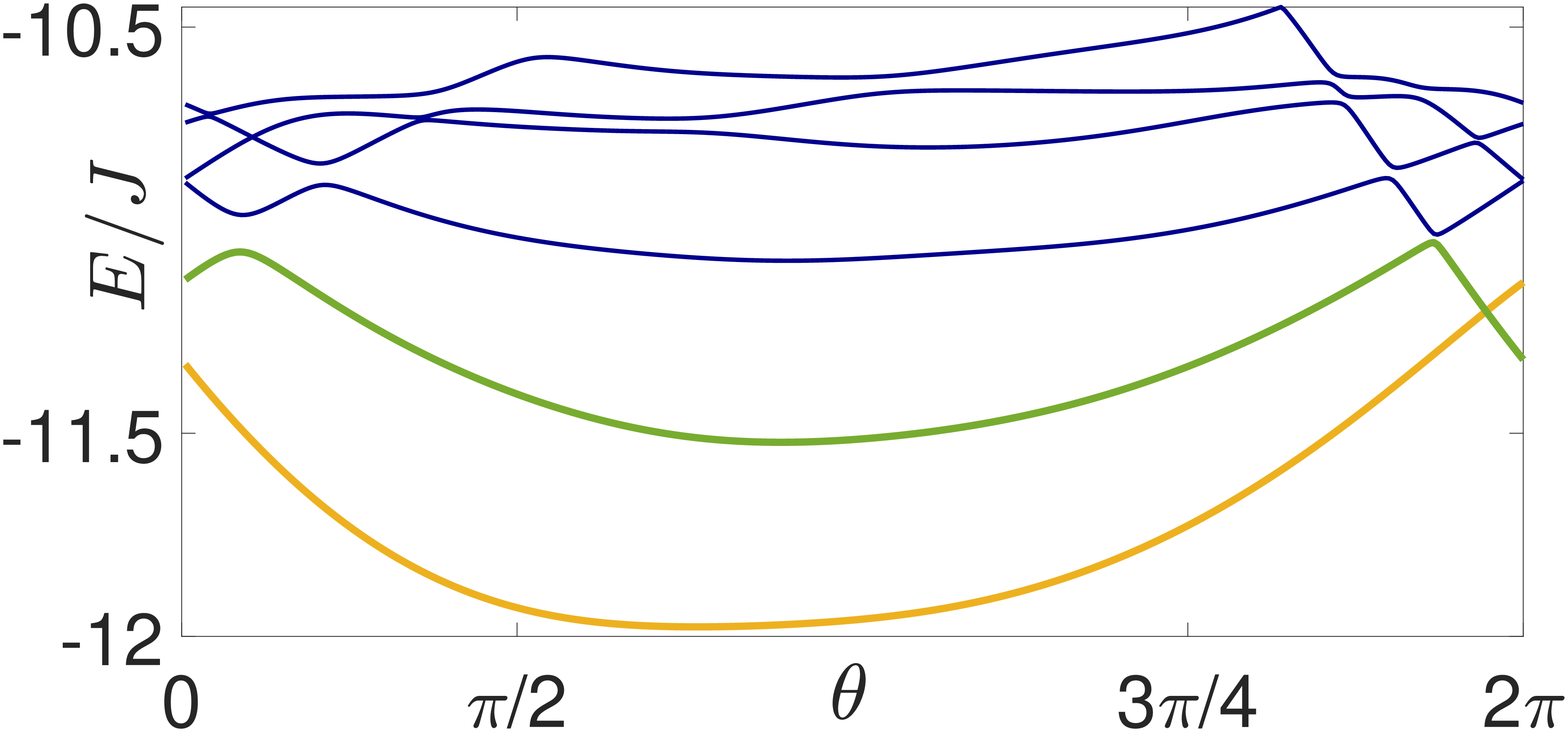}
	\end{minipage}
	\begin{minipage}{0.32\textwidth}
		\begin{picture}(0,0)
		\put(-55,2){{\large (c)}}
		\end{picture}
		\includegraphics[width=\linewidth, height=\textheight,keepaspectratio]{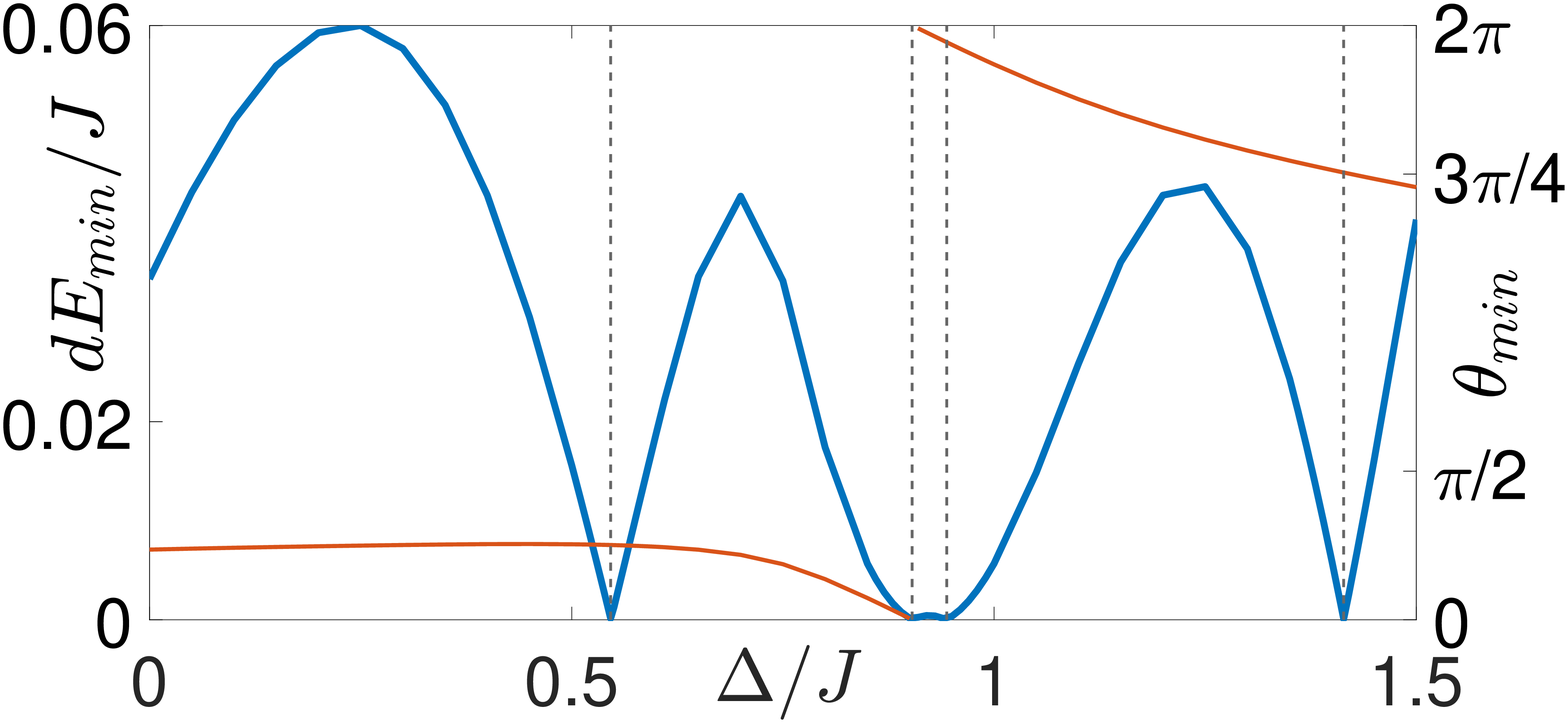}
	\end{minipage}
	\caption{Adiabatic evolution of the low energy eigenstates of the $3\times3$ quantum spin lattice under a rotation of the magnetization of one edge. (a) Adiabatic evolution for $D=J$ and $\Delta=J$. As for most interactions, the ground state does not exhibit an energy crossing with energetically higher states. Thus, no skyrmion is created. (b)  Energy evolution for $D=J$, $\Delta\approx0.943$. The gap closes and a quantum skyrmion is created at $\theta=2\pi$. (c) Evolution of the minimal energy gap $dE_{min}$ between the ground state and the first excited state during the rotation over the exchange anisotropy $\Delta$ and for $D=J$. We find a gap closure for four different values of $\Delta$.  
	}
	\label{OneEdgeRotation}
\end{figure*}
\begin{figure*}[!tb]
	\centering
	\includegraphics[width=0.45\textwidth,trim=70 0 0 0,clip,keepaspectratio]{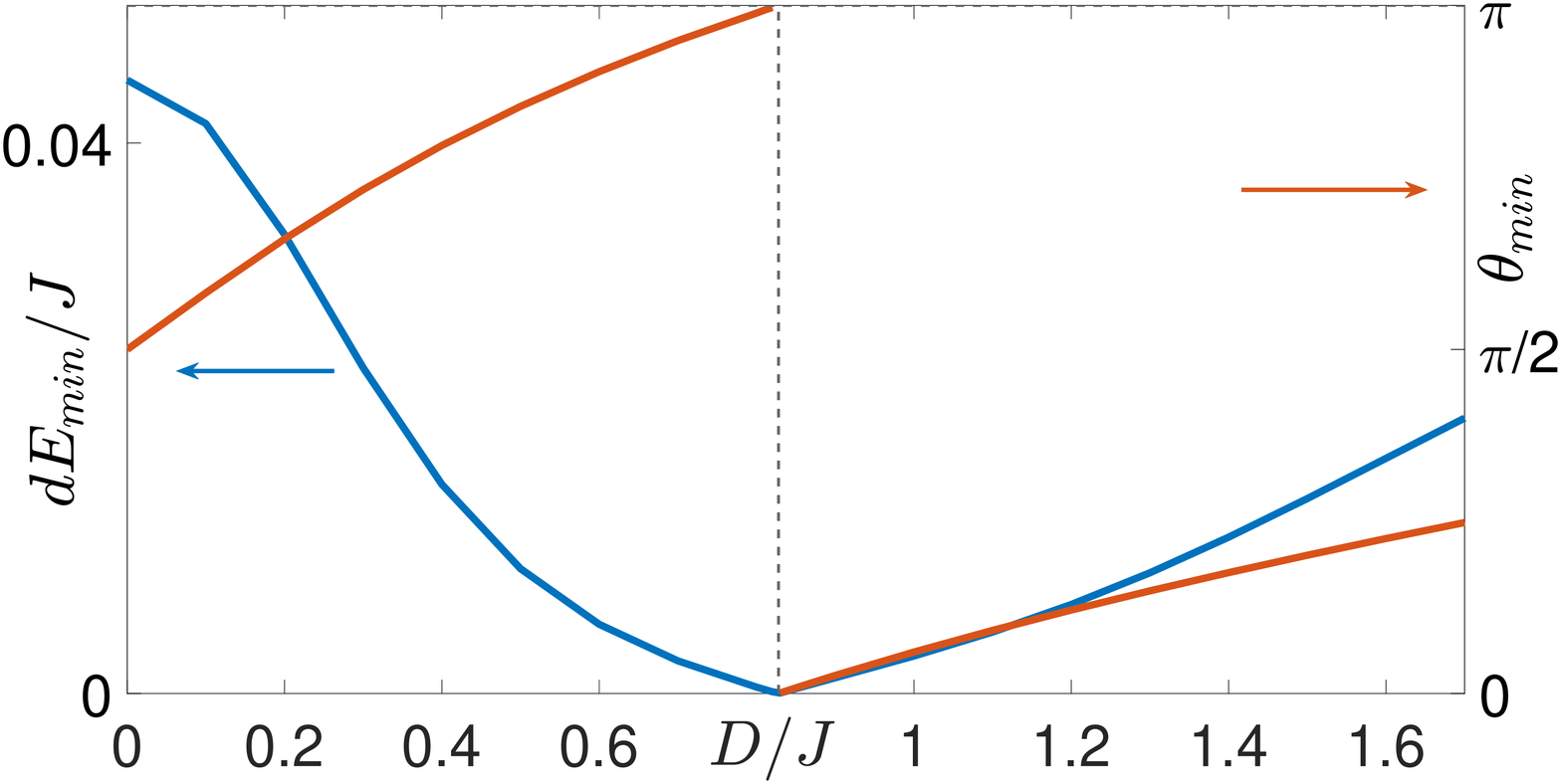}	
	\caption{Minimal energy gap  $d E_{\text{min}}$ between the ground state and higher lying states and the corresponding angle $\theta_{min}$ at which it appears. We consider the $4\times4$ lattice with isotropic exchange ($\Delta=J$) for different DMI strengths $D$. The energy gap closes at only one value of the interaction parameter $D\approx0.82J$ at $\theta=0$ and $\theta=\pi$. Thus, it is not a consequence of the rotation but a degeneracy caused by the specific combination of interactions.}
	\label{fig:Fig3}
\end{figure*}
We show in the main text that a quantum skyrmion can be created in a $3\times3$ quantum spin lattice for a range of interaction parameters if the magnetization at all boundaries is rotated via
\begin{equation}\label{AdiabRot2}
\textbf{S}_{i}=\left(\pm\frac{1}{2}\sqrt{w_{i}}\sin(\theta),\pm\frac{1}{2}\sqrt{1-w_{i}}\sin(\theta),-\frac{1}{2}\cos(\theta)\right).
\end{equation}
The weights $w_i$ we use for the simulation depend on the position $i$ at the boundary. 
For the edge where $x=0$ or $x=4$ and $y$ from $0$ to $4$, the corresponding weights are $W=(w_0,w_1,w_2,w_3,w_4)=(1/2,2/3,1,2/3,1/2)$ corresponding to the $y$ position.
For the edge where $y=0$ or $y=4$ and $x$ runs from $0$ to $4$ the corresponding weights are $W=(1/2,1/3,0,1/3,1/2)$ corresponding to the $x$ position. This corresponds to an exchange of the $x$ and $y$ components of the edge magnetic moments in dependence on the edge.
The given choice of weights is not the only one to create a quantum skyrmion.
Up to our knowledge, all rotation schemes that fulfill the following prerequisites induce the necessary energy level crossings in the $3\times3$ lattice.
First, the rotated boundary magnetic moments need to have no $z-$component of the magnetization at the angles $\theta=\pi/2$ and $\theta=3\pi/4$. This condition is met by the class of rotations of four edges described by Eq.~(\ref{AdiabRot2}).
Second, the rotation needs to be $C_{4\nu}$-symmetric, regarding a rotation around the $z$-axis, i.e., the boundary magnetization is invariant under a rotation  by $\pi/2$ around the $z$-axis. 
Thus, also, e.g., a uniform rotation with $W=(w_0,w_1,w_2,w_3,w_4)=(1,1,1,1,1)$ at the edge with $x=0$ or $x=4$ and $W=(0,0,0,0,0)$ for the edge with $y=0$ or $y=4$, leads to the same qualitative results as the symmetric rotation and a quantum skyrmion can be created after half a rotation for appropriate interaction parameters.\\
The weights we used for the boundary rotation for the $4\times4$ lattice are $W=(w_0,w_1,w_2,w_3,w_4,w_5,w_6)=(0.5,0.6,\allowbreak 0.8, \allowbreak 0.8,0.6,0.5)$ for the edge at $x=0$ or $x=6$ and $W=(0.5,0.4,0.2,0.2,0.4,0.5)$ for the edge at $y=0$ or $y=6$. Also for the $4\times4$ lattice, different weights that fulfill the necessary prerequisites introduced for the $3\times3$ lattice, lead to the same qualitative results as discussed in the main text and in section \ref{4x4Creation} of the Appendix.\\
\textit{One-Edge Rotation.--}
A classical magnetic skyrmion can be created starting from a ferromagnetic initial state by a rotation of the boundary magnetization at only one edge, see Ref. \cite{Schaffer2020}.  
In a quantum spin lattice, in contrast, we find that quantum skyrmions can in general not be created by an adiabatic rotation of only one edge. 
We rotate the magnetization of one edge according to Eq.~(\ref{AdiabRot2}) and with $W=(1/2,1/3,0,1/3,1/2)$. If only one edge is rotated, neither of the two conditions to induce a level crossing is met.
In Fig.~\ref{OneEdgeRotation}~(a), the energy evolution during the one-edge rotation for $D=\Delta=J$ is depicted. 
No level-crossings occur, i.e., no quantum skyrmion state is reached.
Only for special values of the exchange anisotropy $\Delta$, the energy gap between the ground and the first excited state closes, as shown in Fig.~\ref{OneEdgeRotation} (b) and (c). Only with an instantaneously closed energy gap to the ground state a change in the spin configuration can be obtained after the rotation. The two energetically lowest states remain separated from the rest of the spectrum.
For $D=J,~\Delta=0.943J$, the rotation creates a quantum skyrmion. For the other parameters that induce an energy level crossing, a change between two different states with equal $C$ occurs. 
Other rotation schemes of the magnetization of one edge show the same qualitative behavior. The values of $\Delta$ at which the energy gap closure appears vary slightly for different rotation schemes.
\section{Appendix: Quantum skyrmion creation in the 4x4 quantum spin lattice}\label{4x4Creation}
We numerically find that creating quantum skyrmions in the $4\times4$ lattice with isotropic exchange ($\Delta=J$) is impossible with the presented symmetric rotation scheme. For $D=0$, the ground state is not degenerate at the rotation angles $\theta=\pi/2$ and $\theta=3\pi/4$.
There is no crossing of the ground state with higher lying energy levels for nearly all studied values of $D$, as shown in Fig.~\ref{fig:Fig3}. The only exception is $D\approx0.82J$, which induces an accidental degeneracy of the two energetically lowest states already prior to the rotation.
In contrast to the $3\times3$ lattice, no quantum skyrmion can be created with the presented rotation scheme for isotropic exchange. 
\end{document}